%% file: 2202_12483.tex
\documentclass[
reprint,
superscriptaddress,
amsmath,amssymb,
aps,
prl,
]{revtex4-2}

\usepackage{graphicx}
\usepackage{dcolumn}
\usepackage{bm}
\usepackage{hyperref}
\usepackage{float}
\newcommand{\SMSecmSummation}{S1} 
\newcommand{\SMSecPruning}{S2} 
\newcommand{\SMSecValidation}{S3} 
\newcommand{\SMSecNegEneEla}{S4} 
\newcommand{\SMSeckUTURelation}{S5} 
\newcommand{\SMSeckSkUExactForms}{S6} 
\newcommand{\SMSecTUinfExactForm}{S7} 
\newcommand{\SMSecDataCollapse}{S8} 
\newcommand{\SMSecMeanFreePath}{S9} 
\newcommand{\SMSecSAWandNAWlimits}{S10} 
\newcommand{\SMSecWList}{S11} 
\newcommand{\SMSecWn}{S12} 
\newcommand{\SMSeckUnegative}{S13} 
\newcommand{\SMSecTUinfList}{S14} 
\newcommand{\SMSecTUinfRationalFunc}{S15} 

\begin{document}

\title{
Solvent-Induced Negative Energetic Elasticity in a Lattice Polymer Chain
}

\author{Nobu C.~Shirai}
\thanks{These authors contributed equally}
\email[Corresponding author. ]{shirai@cc.mie-u.ac.jp}
\affiliation{
Center for Information Technologies and Networks, Mie University, Tsu, Mie 514-8507, Japan
}
\author{Naoyuki Sakumichi}
\thanks{These authors contributed equally}
\email[Corresponding author. ]{sakumichi@tetrapod.t.u-tokyo.ac.jp}
\affiliation{
Graduate School of Engineering, The University of Tokyo, 7-3-1 Hongo, Bunkyo-ku, Tokyo 113-8656, Japan
}

\date{\today}

\begin{abstract}
The negative internal energetic contribution to the elastic modulus (negative energetic elasticity) has been recently observed in polymer gels.
This finding challenges the conventional notion that the elastic moduli of rubberlike materials are determined mainly by entropic elasticity.
However, the microscopic origin of negative energetic elasticity has not yet been clarified.
Here, we consider the $n$-step interacting self-avoiding walk on a cubic lattice as a model of a single polymer chain (a subchain of a network in a polymer gel) in a solvent.
We theoretically demonstrate the emergence of negative energetic elasticity based on an exact enumeration up to $n=20$ and analytic expressions for arbitrary $n$ in special cases.
Furthermore, we demonstrate that the negative energetic elasticity of this model originates from the attractive polymer--solvent interaction, which locally stiffens the chain and conversely softens the stiffness of the entire chain. 
This model qualitatively reproduces the temperature dependence of negative energetic elasticity observed in the polymer-gel experiments, indicating that the analysis of a single chain can explain the properties of negative energetic elasticity in polymer gels.
\end{abstract}

\maketitle

Since the widespread acceptance of the macromolecular hypothesis~\cite{Staudinger1920}, the entropic elasticity originating from flexible polymer chains in rubberlike materials has been investigated experimentally and theoretically~\cite{MeyerFerri1935,AnthonyGuth1942,Mark1976}. 
The simplest theoretical explanations for entropic elasticity are provided by statistical models of ideal chains.
For example, the random walk, freely jointed chain, and freely rotating chain models are described in textbooks on statistical mechanics~\cite{Kittel1969book,Callen1991book}, polymer physics~\cite{deGennes1979book,RubinsteinColby2003book}, and soft matter physics~\cite{Doi2013book}.

\begin{figure}[b!] 
\centering
\includegraphics[width=0.99\linewidth]{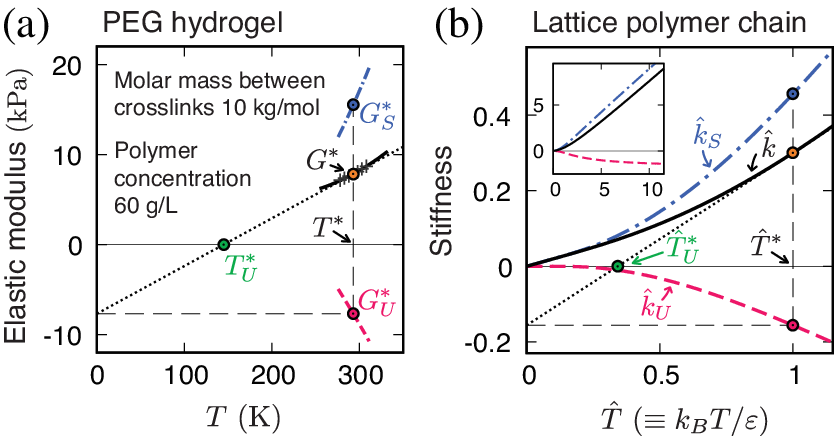}
\caption{\label{fig:G_and_k}Experimental results for a polymer gel~\cite{YoshikawaSakai2021,FujiyabuSakumichi2021} and theoretical results of the lattice polymer chain model.
(a) Temperature dependence of shear modulus $G$ of poly(ethylene glycol) (PEG) hydrogel for $T=278$--$308$~$\mathrm{K}$, including the midpoint $T^*=293$~$\mathrm{K}$ [seven points, including the orange point ($G^*$) in the center].
These data are obtained from Refs.~\cite{YoshikawaSakai2021,FujiyabuSakumichi2021}.
The black solid curve is the fitted quadratic function of these points.
The dotted line is the tangent of the solid curve at the reference temperature $T^*$, which intersects with the horizontal axis at $T_U^*$~($>0$) and the vertical axis at $G_U^*$~($<0$).
Temperature dependencies of $G_U$ (pink dashed curve) and $G_S$ (blue dot-dashed curve) are calculated from the fitted quadratic function.
(b) Temperature dependence of stiffness ($\hat{k}$) of lattice polymer chain model for $(n,r)=(20,10a)$ and its energetic ($\hat{k}_U$) and entropic ($\hat{k}_S$) contributions.
The tangent (dotted line) of $\hat{k}$ at $\hat{T}^*$ intersects with the horizontal axis at $\hat{T}_U^*$~($>0$) and vertical axis at $\hat{k}_U^*\equiv \hat{k}_U(r,\hat{T}^*)$~($<0$).
}
\end{figure}

Rubberlike materials composed of polymer chains exhibit an interplay between entropic ($G_S$) and energetic ($G_U$) contributions to the elastic modulus ($G=G_S+G_U$).
In the case of conventional natural and synthetic rubbers, $\left|G_U\right|$ is significantly smaller than $G_S$~\cite{MeyerFerri1935,AnthonyGuth1942,Mark1976}.
A small $\left|G_U\right|$ is considered to originate from the conformational change of polymer chains, which has been theoretically modeled such as the rotational isometric state model~\cite{Flory1969book}.
By contrast, the significant negative $G_U$ was recently observed in a chemically crosslinked polymer gel, i.e., a polymer network containing a large amount of solvent~\cite{YoshikawaSakai2021,SakumichiSakai2021,FujiyabuSakumichi2021}.
In this observation, $\left|G_U\right|$ is significantly larger than that of the energetic elasticity originating from conformational changes, and $\left|G_U\right|$ reaches the same order of magnitude as $G$, as shown in Fig.~\ref{fig:G_and_k}(a).

Although two previous studies~\cite{YoshikawaSakai2021,FujiyabuSakumichi2021} measured the shear modulus $G$ over a narrow temperature ($T$) range where $G$ can be approximated as a linear function of $T$, the combination of both results shows that $G$ is an increasing convex function of $T$, as shown in Fig.~\ref{fig:G_and_k}(a).
Here, $T$ dependence of $G$ is fitted by quadratic function (black solid curve), and the tangent of the fitted curve (dotted line) at the reference temperature $T^*=293$~$\mathrm{K}$ intersects with the horizontal axis at $T_U^*$ and vertical axis at $G_U^*$.
These two experiments had the same conditions, excluding the temperature ranges; Ref.~\cite{YoshikawaSakai2021} used $T=278$--$298$~$\mathrm{K}$, whereas Ref.~\cite{FujiyabuSakumichi2021} used $T=288$--$308$~$\mathrm{K}$.
By shifting $T^*$ from $288$ (Ref.~\cite{YoshikawaSakai2021}) to $298$~$\mathrm{K}$ (Ref.~\cite{FujiyabuSakumichi2021}), the values of $T_U^*$ increased.
These results imply that (i) $G$ is an increasing convex function of $T$, (ii) $T_U^*$ depends on the reference temperature $T^*$, and (iii) $G_U$ is a monotonically decreasing function of $T$. 

In this Letter, focusing on a subchain (i.e., a chain between adjacent crosslinks) in the polymer network of polymer gels, we theoretically demonstrate the emergence of negative energetic elasticity in a single polymer chain model on a lattice.
Notably, the previous experimental studies~\cite{YoshikawaSakai2021,FujiyabuSakumichi2021} show that $G$ is described as $G=a(T-T_U^*)$ in a narrow temperature range, where $a$ depends on the polymer network topology, but $T_U^*$ does not [see Eqs.~(7) and (9) in Ref.~\cite{SakumichiSakai2021}].
Thus, network topology does not contribute to the proportion of energetic elasticity to entropic elasticity ($G_U^*/G_S^*=-T_U^*/T^*$), suggesting that a subchain is sufficient to investigate $G_U^*/G_S^*$.
As shown in Figs.~\ref{fig:G_and_k}(a) and \ref{fig:G_and_k}(b), this polymer chain model explains $G_U$ as a monotonically decreasing function of temperature in polymer gel experiments.
Furthermore, we provide a microscopic mechanism for the emergence of negative energetic elasticity by examining the polymer--solvent interaction strength in this model.

\begin{figure}[t!] 
\centering
\includegraphics[width=0.98\linewidth]{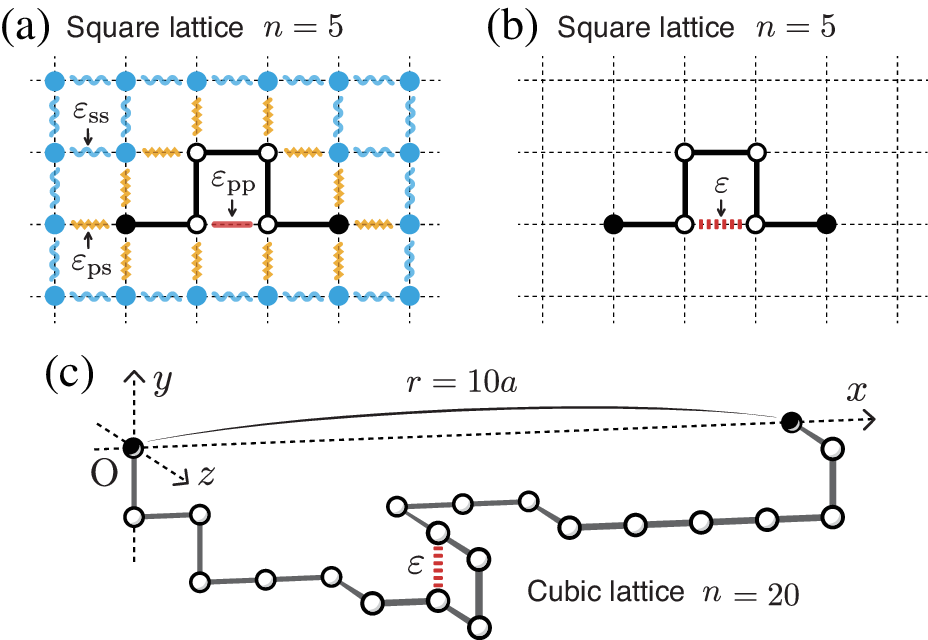}
\caption{\label{fig:interaction}
(a) Two-component model of single polymer chain (open and filled black circles connected with lines) and solvent molecules (light-blue filled circles) on square lattice.
There are three types of nearest-neighbor interactions. 
(b) Reduced model mathematically equivalent to (a).
(c) Example of interacting self-avoiding walk on cubic lattice with on-axis constraint on end-to-end vector. 
}
\end{figure}

\textit{Lattice polymer chain model.}---We consider a single polymer chain model surrounded by solvent molecules on a simple cubic lattice (three dimensions).
This model was first introduced by Orr for highly dilute polymer solutions~\cite{Orr1947} and is one of the simplest ways to express the interaction between a polymer chain and solvent molecules.
In Figs.~\ref{fig:interaction}(a) and \ref{fig:interaction}(b), we use a square lattice (two dimensions) for illustration.
As indicated in Fig.~\ref{fig:interaction}(a), this model consists of solvent molecules and a polymer chain represented by an $n$-step self-avoiding walk (SAW)~\cite{MadrasSlade1996book}, which has $n$ bonds connecting $n+1$ consecutive and distinct lattice sites (i.e., the polymer segments). 
The energy function of the model is given by
\begin{equation} 
E(\omega) = \varepsilon_\mathrm{pp} m_\mathrm{pp} (\omega) + \varepsilon_\mathrm{ps} m_\mathrm{ps}(\omega) + \varepsilon_\mathrm{ss} m_\mathrm{ss}(\omega), \label{eq:Etot}
\end{equation}
where $\omega$ denotes the configuration of SAWs, and $m_\mathrm{pp}(\omega)$, $m_\mathrm{ps}(\omega)$, and $m_\mathrm{ss}(\omega)$ are the numbers of the polymer--polymer, polymer--solvent, and solvent--solvent contact pairs, respectively.
Here, the interaction energies acting between each pair are $\varepsilon_\mathrm{pp}$, $\varepsilon_\mathrm{ps}$, and $\varepsilon_\mathrm{ss}$, respectively. 
In Eq.~(\ref{eq:Etot}), we do not consider the bending energetic terms, which are essential for semiflexible polymers~\cite{BroederszMacKintosh2014}, to focus on the solvent-induced energetic elasticity.

Once the entire lattice size is given, the total number of contact pairs, $m_\mathrm{pp}(\omega)+m_\mathrm{ps}(\omega) + m_\mathrm{ss}(\omega)$, is constant.
In addition, we derive $2m_\mathrm{pp}(\omega)+m_\mathrm{ps}(\omega)=(z-2)n+z$ by counting solvent molecules surrounding $\omega$.
Here, $z$ is the coordination number of a lattice, e.g., $z=4$ and $6$ for the square and cubic lattices, respectively.
Thus, when $n$ is constant, Eq.~(\ref{eq:Etot}) is rewritten~\cite{Orr1947,Doi2013book} as
\begin{equation} 
E(\omega) = \varepsilon\, m(\omega), \label{eq:E}
\end{equation}
where a constant term has been omitted, and $\varepsilon \equiv \varepsilon_\mathrm{pp} - 2 \varepsilon_\mathrm{ps} + \varepsilon_\mathrm{ss}$ and $m(\omega)\equiv m_\mathrm{pp}(\omega)$. 
Here, the original model illustrated in Fig.~\ref{fig:interaction}(a) was reduced to the so-called interacting SAW~\cite{Vanderzande1998book} shown in Fig.~\ref{fig:interaction}(b), which is a single-chain system with the intrachain interaction.
The interacting SAW reduces to the (noninteracting) SAW at $\varepsilon=0$.
Notably, many studies on the interacting SAW have focused on the self-attractive condition $(\varepsilon<0)$ to investigate the collapsing transition~\cite{Rapaport1974,MarenduzzoSeno2003,KumarGuttmann2007,LeeLee2012,HsiehHu2016aug}. 
By contrast, this study focuses mainly on the self-repulsive condition ($\varepsilon>0$) to investigate the effect of attractive polymer--solvent interactions.

\textit{Energetic and entropic elasticities in lattice polymer chain.}---To calculate the stiffness (a single-chain counterpart of the elastic modulus) of the lattice polymer chain model, we impose an on-axis constraint on the end-to-end vector of $\omega$.
Here, the length of the vector is $r$, and the direction of the vector is the same as the $x$ axis, as shown in Fig.~\ref{fig:interaction}(c). 
The partition function with the on-axis constraint using Eq.~(\ref{eq:E}) is given by
\begin{equation} 
Z(r,T) = \sum_{m=0}^{m_\mathrm{ub}} W_{n,m}(r)\,e^{- \varepsilon m/(k_B T)}, \label{eq:Z}
\end{equation}
where $k_B$ is the Boltzmann constant, $T$ is the absolute temperature, and $W_{n,m}(r)$ is the number of possible $\omega$ for a given set of $n$, $r$, and $m$.
In Eq.~(\ref{eq:Z}), $m_\mathrm{ub}$ is an upper bound of $m$ summation [i.e., $W_{n,m}(r)=0$ for $m\geq m_\mathrm{ub}+1$], which is discussed in Supplemental Material, Sec.~\SMSecmSummation{} ~\cite{SMcommentGelSAW2021}. 
The corresponding free energy is $A(r,T) = - k_B T \ln Z(r,T)$.

We define the stiffness of the lattice polymer chain model with the on-axis constraint.
In the continuum limit ($n\to\infty$ and lattice spacing $a\to 0$), the stiffness is defined as the second derivative of the free energy: 
\begin{equation}
k(r,T)\equiv \frac{\partial^2 A(r,T)}{\partial r^2}=k_B T\left[ \left( \frac{\frac{\partial Z(r,T)}{\partial r}}{Z(r,T)} \right)^2 
- \frac{\frac{\partial^2 Z(r,T)}{\partial r^2}}{Z(r,T)} \right].
\end{equation}
Thus, we define the finite difference form of stiffness as
\begin{eqnarray}
k(r,T) &\equiv& k_B T\left[ \left( \frac{1}{Z(r,T)}\sum_{m=0}^{m_\mathrm{ub}} \frac{\varDelta W_{n,m}(r)}{\varDelta r} \,e^{-\varepsilon m/(k_B T)} \right)^2 \right.\notag\\
&& \left. \quad - \frac{1}{Z(r,T)}
\sum_{m=0}^{m_\mathrm{ub}} \frac{\varDelta^2 W_{n,m}(r)}{\varDelta r^2} \,e^{- \varepsilon m/(k_B T)}
 \right], 
\label{eq:k}
\end{eqnarray}
where the first- and second-order differences of $W_{n,m}(r)$ are given by
$\varDelta W_{n,m}(r) \equiv \left[W_{n,m}(r+\varDelta r)- W_{n,m}(r-\varDelta r)\right]/2$, 
and
$\varDelta^2 W_{n,m}(r) \equiv W_{n,m}(r+\varDelta r) - 2W_{n,m}(r)+ W_{n,m}(r-\varDelta r)$, 
respectively.
Here, $\varDelta r \equiv 2a$ because $\omega$ exists only for odd $\hat{r}\equiv r/a$ for odd $n$ and only for even $\hat{r}$ for even $n$ (see Supplemental Material, Sec.~\SMSecWList{}~\cite{SMcommentGelSAW2021}). 

We decompose the elasticity into its energetic and entropic contributions as $k=k_U+k_S$ in the same way as in Refs.~\cite{YoshikawaSakai2021,SakumichiSakai2021}. 
According to thermodynamics, $A=U-TS$, where $U$ is the internal energy, and $S$ is the entropy.
Thus, in the continuum limit, the energetic and entropic contributions are $k_U(r,T)\equiv \partial^2 U(r,T)/\partial r^2$ and $k_S(r,T)\equiv -T \partial^2 S(r)/\partial r^2$, respectively.
From Maxwell's relation, we have $k_S(r,T)= T\partial k(r,T)/\partial T$.
Thus, we calculate $k_S(r,T)= T\partial k(r,T)/\partial T$ and $k_U=k-k_S$ in the lattice polymer chain model using Eq.~(\ref{eq:k}).

\textit{Exact enumeration and derivation of polynomial functions.}---We exactly enumerate $W_{n,m}(r)$ for $1\leq n\leq 20$ using the simplest recursive algorithm~\cite{Martin1962} with two pruning algorithms, considering the octahedral symmetry of the simple cubic lattice and the reachability of $\omega$ to a specific endpoint (see Supplemental Material, Sec.~\SMSecPruning{}~\cite{SMcommentGelSAW2021}). 
The complete lists of $W_{n,m}(r)$ are provided in Sec.~\SMSecWList{} of Supplemental Material~\cite{SMcommentGelSAW2021}. 
Those lists are consistent with the results reported in Refs.~\cite{Orr1947,FisherSykes1959,Sykes1961,Sykes1963,FisherHiley1961,NemirovskyDouglas1992,ButeraComi1999,Guttmann1987,DombWilmers1965,A001412,A174319} (see Supplemental Material, Sec.~\SMSecValidation{}~\cite{SMcommentGelSAW2021}). 
Using the lists of $W_{n,m}(r)$ with additional results up to $n=26$ for $r=(n-8)a$, we exactly derive the polynomial functions $W_{n,m}\big((n-2)a\big)$, $W_{n,m}\big((n-4)a\big)$, $W_{n,m}\big((n-6)a\big)$, and $W_{n,m}\big((n-8)a\big)$ for arbitrary integers $n$ and $m$, which are provided in Sec.~\SMSecWn{} of SM~\cite{SMcommentGelSAW2021}. 

\textit{Emergence of negative energetic elasticity.}---From the lists of $W_{n,m}(r)$, we can analytically calculate $k$, $k_U$, and $k_S$.
Figure~\ref{fig:G_and_k}(b) shows a representative result for $\varepsilon>0$ and $(n,r)=(20,10a)$.
Here, we introduce the dimensionless quantities
$\hat{k}\equiv a^2 k/\varepsilon$, 
$\hat{k}_U\equiv a^2 k_U /\varepsilon$, 
$\hat{k}_S\equiv a^2 k_S/\varepsilon$, and
$\hat{T}\equiv k_B T/\varepsilon$.
Figure~\ref{fig:G_and_k}(b) demonstrates the emergence of the solvent-induced negative energetic elasticity ($k_U<0$) in the model.
Figure~\ref{fig:interaction}(c) displays an example of $\omega$ for $r=10a$.
In this Letter, we use $r=10a$ for the illustration [e.g., Fig.~\ref{fig:G_and_k}(b)] because the maximum value of $|k_U|/k$ is larger than $r\neq 10a$. 
Although the extent of $k_U/k$ depends on $r$, the negative $k_U$ can be observed for different $n\geq 6$ and $r$. 
Notably, we find that $k_U<0$ for arbitrary $n\geq 13$, $\varepsilon>0$, and positive finite $T$ using the polynomial functions $W_{n,m}\big((n-2)a\big)$, $W_{n,m}\big((n-4)a\big)$, $W_{n,m}\big((n-6)a\big)$, and $W_{n,m}\big((n-8)a\big)$ (see Supplemental Material, Sec.~\SMSeckUnegative{}~\cite{SMcommentGelSAW2021}).

Figures~\ref{fig:G_and_k}(a) and (b) demonstrate the qualitative consistency between the previous experimental results for the shear modulus $G$ of the poly(ethylene glycol) (PEG) hydrogel~\cite{YoshikawaSakai2021,FujiyabuSakumichi2021} and our results.
These results suggest that negative energetic elasticity ($G_U<0$) in the polymer gel originates from a single chain ($k_U<0$).

\begin{figure}[H] 
\centering
\includegraphics[width=0.99\linewidth]{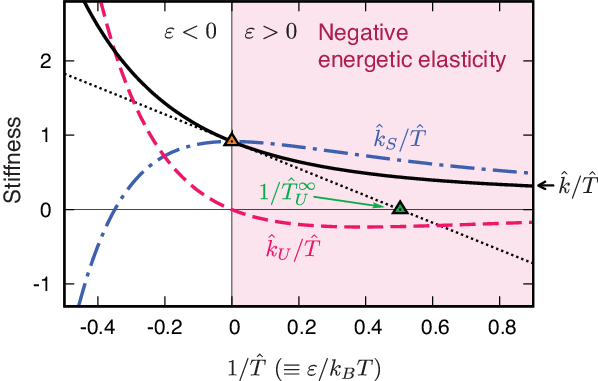}
\caption{\label{fig:k_over_T}
Analytic curves of $\hat{k}/\hat{T}$, $\hat{k}_U/\hat{T}$, and $\hat{k}_S/\hat{T}$ as functions of $1/\hat{T}$ for $(n,r)=(20,10a)$. 
The tangent (dotted line) of $\hat{k}/\hat{T}$ at $1/\hat{T}^* = 0$ intersects with the horizontal axis at $1/\hat{T}_U^\infty$. 
The condition for negative energetic elasticity ($\hat{k}_U/\hat{T}<0$) is $\varepsilon>0$.
}
\end{figure}

To examine the effect of the sign of $\varepsilon$ on the energetic elasticity, we show the dependence of the stiffness $\hat{k}/\hat{T}=a^2 k/(k_B T)$ on $1/\hat{T}\equiv \varepsilon/(k_B T)$ in Fig.~\ref{fig:k_over_T}.
[Note that $\hat{k}\to\infty$ and $\hat{k}_S\to\infty$ at $\varepsilon=0$ (noninteracting SAW), whereas $0<\hat{k}/\hat{T}<\infty$ and $0<\hat{k}_S/\hat{T}<\infty$.]
Figure~\ref{fig:k_over_T} depicts that the energetic contributions are negative, zero, and positive for $\varepsilon>0$, $\varepsilon=0$, and $\varepsilon>0$, respectively, for $(n,r)=(20,10a)$.

\hfill

As depicted in Fig.~\ref{fig:G_and_k}(b), $\hat{T}_U^*\equiv\hat{T}_U(\hat{T}^*)$ denotes the $\hat{T}$ intercept of the tangent of $\hat{k}=\hat{k}(\hat{T})$ at the reference temperature $\hat{T}=\hat{T}^*$.
For polymer gels, $T_U^*$ [Fig.~\ref{fig:G_and_k}(a)] is a key factor in the analysis of negative energetic elasticity because $T_U^*$ does not depend on the polymer network topology~\cite{YoshikawaSakai2021,SakumichiSakai2021}.
In addition, in the lattice polymer chain model, $T_U^*$ is a better measure of the negative energetic elasticity than $k_U$, because $T_U^*=\varepsilon\hat{T}_U^*/k_B$ is independent of the lattice spacing $a$, unlike $k_U=\varepsilon\hat{k}_U/a^2$, which depends on $a$.

\begin{figure}[t!] 
\centering
\includegraphics[width=0.99\linewidth]{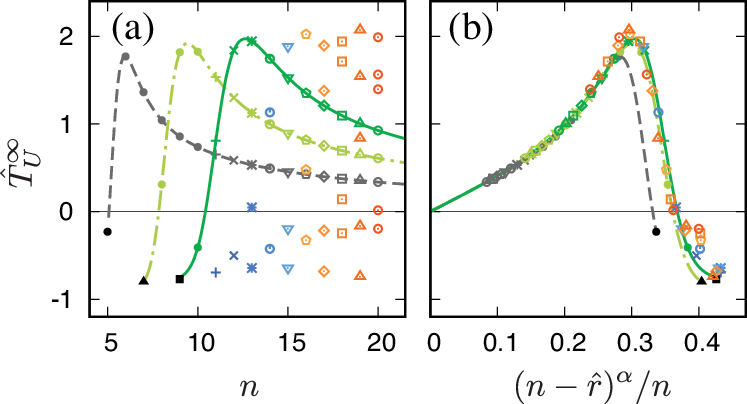}
\caption{\label{fig:T_U_inf_n20}(a) Exact values of $\hat{T}_U^\infty(n,r)$ for $n=5,6,\dots,20$ and three analytic curves for $r=(n-2)a$, $(n-4)a$, and $(n-6)a$, shown as gray dashed, light-green dot-dashed, and green solid curves, respectively. 
(b) Same points and curves are plotted as functions of $(n-\hat{r})^\alpha/n$, where $\alpha=3/4$.
These collapse onto a single master curve.
}
\end{figure}

We define $\hat{T}_U^\infty \equiv\lim_{\hat{T}^*\to \infty}\hat{T}_U(\hat{T}^*)$, which is a good indicator of the negative energetic elasticity in the sense that $\hat{T}_U^\infty>0$ is identical to $\hat{k}_U<0$ in the case of $\varepsilon>0$. 
As shown in Fig.~\ref{fig:k_over_T}, the $1/\hat{T}$ intercept of the tangent of $\hat{k}/\hat{T}$ at $1/\hat{T}=0$ corresponds to $1/\hat{T}_U^\infty$.
Notably, $\hat{T}_U^\infty$ is a functional of $W_{n,m}(r)$ and is a rational number for a given set of $n$, $r$, and $m$ (see Supplemental Material, Secs.~\SMSeckUTURelation{}, \SMSeckSkUExactForms{}, and~\SMSecTUinfExactForm{}~\cite{SMcommentGelSAW2021}).
In Fig.~\ref{fig:T_U_inf_n20}(a), we plot $\hat{T}_U^\infty$ that is calculated from $W_{n,m}(r)$ 
(the exact rational numbers of $\hat{T}_U^\infty$ are listed in Sec.~\SMSecTUinfList{} of Supplemental Material~\cite{SMcommentGelSAW2021}). 

We successfully determine the analytic expressions of $\hat{T}_U^\infty$ as the rational functions of $n$ for $r=(n-2)a$, $(n-4)a$, and $(n-6)a$, using the polynomial functions of $W_{n,m}(r)$.
For example,
\begin{widetext}
\begin{equation}
\label{eq:hat_T_U_inf_rational_func}
\hat{T}_U^\infty\big(n,(n-2)a\big) = \frac{4(15n^8 - 356n^7 + 3766n^6 - 23016n^5 + 88019n^4 - 213804n^3 + 317784n^2 - 256008n + 81484)}{(n - 1)(9n^8 - 204n^7 + 2026n^6 - 11648n^5 + 42733n^4 - 102444n^3 + 156272n^2 - 137656n + 53028)}. 
\end{equation}
\end{widetext}
The other rational functions, $\hat{T}_U^\infty\big(n,(n-4)a\big)$ and $\hat{T}_U^\infty\big(n,(n-6)a\big)$, are provided in Sec.~\SMSecTUinfRationalFunc{} of Supplemental Material~\cite{SMcommentGelSAW2021}. 
In Fig.~\ref{fig:T_U_inf_n20}(a), we overlay the three curves of these functions, which pass through all the corresponding points of $\hat{T}_U^\infty$.

Figure~\ref{fig:T_U_inf_n20}(b) shows $\hat{T}_U^\infty$ as a function of $(n-\hat{r})^\alpha/n$ for $\alpha=3/4$.
Here, the three curves of the analytic expressions and the points collapse onto a single master curve, except for the small values of $n$.
A possible origin of the exponent $\alpha=3/4$ is the reduction of the dimensionality from three to two by the on-axis constraint on the end-to-end vector, resulting in the universal critical exponent $\nu=3/4$ of the two-dimensional SAW~\cite{Nienhuis1982,Nienhuis1984} (see Supplemental Material, Sec.~\SMSecDataCollapse{}~\cite{SMcommentGelSAW2021}).

\textit{Microscopic mechanism of negative energetic elasticity.}---To characterize the microscopic properties of the lattice polymer chain model with a negative $k_U$, we evaluate the thermal average of the mean free path of the interacting SAW as
\begin{equation} 
\ell(r,T) = \frac{1}{Z(r,T)}
\sum_{m=0}^{m_\mathrm{ub}} \sum_{b=0}^{n-1}\frac{na}{b+1}W_{n,m,b}(r)\,e^{- \varepsilon m/(k_B T)},
\label{eq:local_stiffness}
\end{equation}
where $b$ is the number of bending points of $\omega$, and $W_{n,m,b}(r)$ is the number of $\omega$ for a given set of $n$, $r$, $m$, and $b$. 
In Eq.~(\ref{eq:local_stiffness}), $na/(b+1)$ is the mean free path for each $\omega$ (see Supplemental Material, Sec.~\SMSecMeanFreePath{}~\cite{SMcommentGelSAW2021}). 
For example, $b=4$ and $13$ for $\omega$ in Figs.~\ref{fig:interaction}(b) and ~\ref{fig:interaction}(c), respectively.

\begin{figure}[htbp] 
\centering
\includegraphics[width=0.98\linewidth]{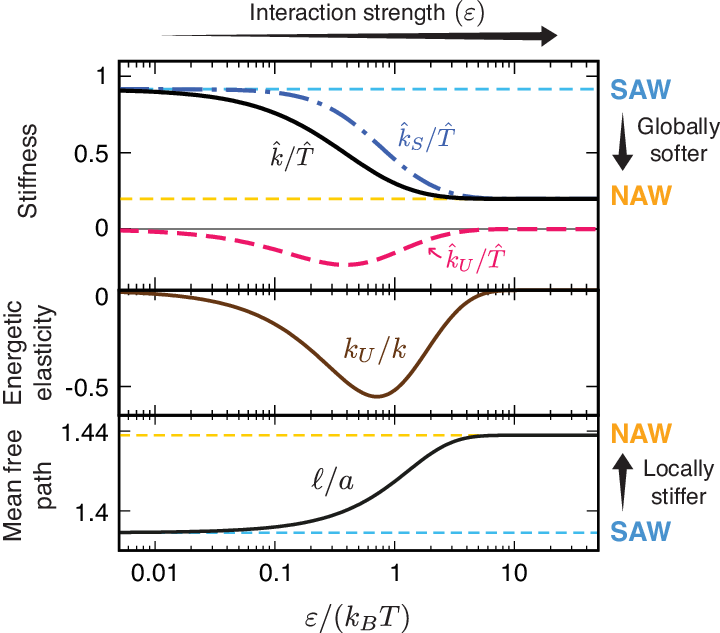}
\caption{\label{fig:pLength_kappa}Polymer--solvent interaction [$\varepsilon/(k_B T)$] dependences of $\hat{k}/\hat{T}$, $\hat{k}_U/\hat{T}$, $\hat{k}_S/\hat{T}$ (top panel), $k_U/k$ (middle panel), and $\ell/a$ (bottom panel) for $(n,r)=(20,10a)$ and $\varepsilon > 0$. 
As $\varepsilon$ increases, the lattice polymer chain model (i.e., interacting SAW) becomes globally softer and locally stiffer.
}
\end{figure}

Figure~\ref{fig:pLength_kappa} indicates that $\hat{k}/\hat{T}=a^2 k/(k_B T)$ increases with $\varepsilon/(k_B T)$, whereas $\ell$ decreases with $\varepsilon/(k_B T)$.
Here, $\hat{k}/\hat{T}$ characterizes the ``global'' stiffness of the whole polymer chain, whereas $\ell$ characterizes the ``local'' stiffness of the chain.
Thus, the global and local stiffnesses are negatively correlated, which is also observed in various sets of $(n,r)$ (see Supplemental Material, Sec.~\SMSecNegEneEla{}~\cite{SMcommentGelSAW2021}).
These results confirm that polymer chains become locally stiffer because of the attractive interaction with solvent molecules, and globally softer because of the smaller curvature of free energy in the case of $\varepsilon>0$ and $\hat{k}_U<0$.
This is the microscopic mechanism of negative energetic elasticity in the lattice polymer chain model.

Figure~\ref{fig:pLength_kappa} also shows values corresponding to the SAW ($\varepsilon \to 0$) and neighbor-avoiding walk (NAW; $\varepsilon \to \infty$)~\cite{TorrieWhittington1975,IshinabeChikahisa1986,NemirovskyDouglas1992} (see Supplemental Material, Sec.~\SMSecSAWandNAWlimits{}~\cite{SMcommentGelSAW2021}). 
The emergence of negative energetic elasticity is characterized by the crossover between the SAW and NAW, which only possess entropic elasticity. 
The minimum of $k_U/k$ is $-0.554$ at $\varepsilon/(k_B T) \simeq 0.712$, where the interaction strength $\varepsilon$ is the same order of magnitude as the thermal energy $k_B T$.

\textit{Concluding remarks.}---We used the simplest lattice polymer chain model to explain both the energetic and entropic elasticities (Fig.~\ref{fig:interaction}).
By exactly enumerating the configurations of this model, we obtained the stiffness of the chain and its energetic and entropic contributions (Fig.~\ref{fig:k_over_T}).
This result demonstrates that the negative energetic elasticity originates from the polymer--solvent interaction.
The three rational functions of $T_U^\infty$ with respect to $n$ are derived from the enumeration results (Fig.~\ref{fig:T_U_inf_n20}), revealing that negative energetic elasticity exists for all finite $n\geq 6$. 
We revealed a negative correlation between the stiffness of the whole polymer chain and the mean free path of the chain (Fig.~\ref{fig:pLength_kappa}).
In short, locally stiffer chains are globally softer.

Although this simple model does not include chemical details, it qualitatively reproduces the temperature dependence of negative energetic elasticity observed in the experiments conducted on the PEG hydrogel~\cite{YoshikawaSakai2021,FujiyabuSakumichi2021} (Fig.~\ref{fig:G_and_k}).
This fact indicates that negative energetic elasticity would emerge in various single polymer chains~\cite{NakajimaNishi2006,KolbergBalzer2019,BaoCui2020} and various polymer gels other than the PEG hydrogel.
Therefore, this model provides a starting point for the further understanding of negative energetic elasticity in polymer chains and networks in solvents.

\vskip.5\baselineskip

\begin{acknowledgments}
We thank Yuki Yoshikawa and Takamasa Sakai for allowing us to use the experimental data in Fig.~\ref{fig:G_and_k}(a) and for their useful comments, especially those regarding the temperature dependence of $T_U^*$. 
This study was supported by JSPS KAKENHI Grant No.~22K13973 to N.C.S., and No.~19K14672 and No.~22H01187 to N.S.
\end{acknowledgments}

\end{document}


\title{
Supplemental Material for:\\
``Solvent-Induced Negative Energetic Elasticity in a Lattice Polymer Chain''
}

\author{Nobu C.~Shirai}
\thanks{These authors contributed equally}
\email[Corresponding author. ]{shirai@cc.mie-u.ac.jp}
\affiliation{%
Center for Information Technologies and Networks, Mie University, Tsu, Mie 514-8507, Japan
}
\author{Naoyuki Sakumichi}%
\thanks{These authors contributed equally}
\email[Corresponding author. ]{sakumichi@tetrapod.t.u-tokyo.ac.jp}
\affiliation{
Graduate School of Engineering, The University of Tokyo, 7-3-1 Hongo, Bunkyo-ku, Tokyo 113-8656, Japan
}

\date{\today}

\maketitle

\section{Upper bound of contact pairs} 
We consider an upper bound of the number of nearest-neighbor contact pairs $m$, which is a non-negative integer, for a configuration $\omega$ of $n$-step SAW.
Notably, the number of segments of the lattice polymer chain is $n+1$.
First, we give a loose upper bound of $m$ from a fully packed $\omega$ without vacant sites inside, which was derived by Orr~\cite{Orr1947}.
The maximum number of contacts for the two segments at both ends of the lattice polymer chain is $z-1$, where $z$ is the coordination number of a lattice, and that for the other $n-1$ segments other than both ends is $z-2$.
Thus, we obtain an upper bound of $m$ as
\begin{eqnarray}
    m^{\mathrm{Orr}}_\mathrm{ub}&=&\frac{2(z-1)+(n-1)(z-2)}{2} \notag\\
    &=&\left(\frac{z}{2}-1\right)n+\frac{z}{2},
\end{eqnarray}
where we divide by the factor two not to doubly count the pairs shared by two segments.

Second, we reduce the upper bound of $m$ by considering segments on the surface of the fully packed $\omega$.
For $n=0$ and $1$, the upper bound $m_\mathrm{ub}=0$ because $\omega$ does not have sufficient length to make a contact.
For $n\geq 2$, the upper bound becomes
\begin{equation}
    m_\mathrm{ub}=m^{\mathrm{Orr}}_\mathrm{ub}-z=\left(\frac{z}{2}-1\right)n-\frac{z}{2},
\end{equation}
because at least one polymer segment exists on the surface of $\omega$ for each direction, which cannot maximize the number of contacts.
In the case of a simple cubic lattice, the upper bound is $m_\mathrm{ub}=2n-3$ because $z=6$.

\section{Two pruning algorithms of enumeration method} 
This section explains the details of the two pruning algorithms of the enumeration method. 

The first algorithm prunes the search tree based on the symmetry of $\omega$ on a lattice (e.g., see Sec.~2.2 of Ref.~\cite{HsiehHu2016aug}).
We select a single configuration from the set of $\omega$, which is generated by symmetry operations including rotation and mirroring.
We multiply the symmetry factor of the selected $\omega$ by its number and prune the search tree for the other members of the set.

The second algorithm prunes the search tree based on the reachability of $\omega$ to a given endpoint.
This algorithm requires the specification of an endpoint for $\omega$. 
To calculate $W_n(r)$ or $W_{n,m}(r)$, we specify the endpoint at $(r,0,0)$.
We calculate the shortest path length to the endpoint and compare it with the number of remaining steps of the recursion.
When the shortest path length exceeds the number of remaining steps, we terminate further search for the branch.

Notably, the enumeration data might be extended using the enumeration techniques developed during the last couple of decades for the (noninteracting) SAW~\cite{SchramBisseling2011,SchramBisseling2013,ClisbySlade2007}.
However, those techniques were never extended to the interacting SAW, and additional efforts are required to store contact information~\cite{HsiehHu2016aug}. 

\section{Validation of enumeration results} 
This section explains the validation of the enumeration results.
We validate the result of $W_n(a)$ in Tables~\ref{tab:W_n_r_n1to19} and \ref{tab:W_n_r_n2to20} using data in Ref.~\cite{ButeraComi1999} up to $n=20$, and validate the result of $W_{13}(r)$ for $r=a,3a,\dots,13a$ in Table~\ref{tab:W_n_r_n1to19} using data in Ref.~\cite{DombWilmers1965}.

To indirectly validate the result of $W_{n,m}(r)$, we also use the results for the number of $n$-step SAWs (without the constraint on the end-to-end vector), $c_n$, and the number of these walks for a given number of $m$, $c_n(m)$, which can be calculated using only the first algorithm. 
Here, $c_n$ and $c_n(m)$ satisfy
\begin{equation}
c_n=\sum_{m=0}^{m_\mathrm{ub}} c_n(m).
\end{equation}
The values of $c_n$ that we obtained are validated by Ref.~\cite{Orr1947} up to $n=6$, by Ref.~\cite{FisherSykes1959} up to $n=9$, by Ref.~\cite{Sykes1961} up to $n=10$, by Ref.~\cite{Sykes1963} up to $n=14$, and by Refs.~\cite{Guttmann1987,A001412} up to $n=18$.
Notably, the values of $c_{11}$ in Ref.~\cite{Sykes1961} and $c_{16}$ in Ref.~\cite{Sykes1963} are inconsistent with the ones in Refs.~\cite{Guttmann1987,A001412}, which are latest and more reliable.
The values of $c_n(m)$ obtained in this study were validated by the numbers in Ref.~\cite{Orr1947} up to $n=6$, those of $c_n(0)$ by Ref.~\cite{FisherHiley1961} up to $n=9$, and those of $c_n(m)$ for $m\leq 5$ by Ref.~\cite{NemirovskyDouglas1992} up to $n=11$. 
Notably, the value of $c_{10}(0)$ in Ref.~\cite{FisherHiley1961} is inconsistent with those reported in Refs.~\cite{NemirovskyDouglas1992,A174319}, which are latest and more reliable.

\section{Polymer--solvent interaction dependencies of stiffness and energetic elasticity\label{sec:n_r_dependencies_of_neg_ene_ela}} 
This section describes the polymer--solvent interaction ($\varepsilon$) dependencies of $\hat{k}/\hat{T}$, $k_U/k$, and $\ell$ for $\varepsilon>0$ and various lengths $n$.
Here, we denote $\hat{T}\equiv k_B T/\varepsilon$.

\begin{figure*}[h!] 
\centering
\includegraphics[width=0.83\linewidth]{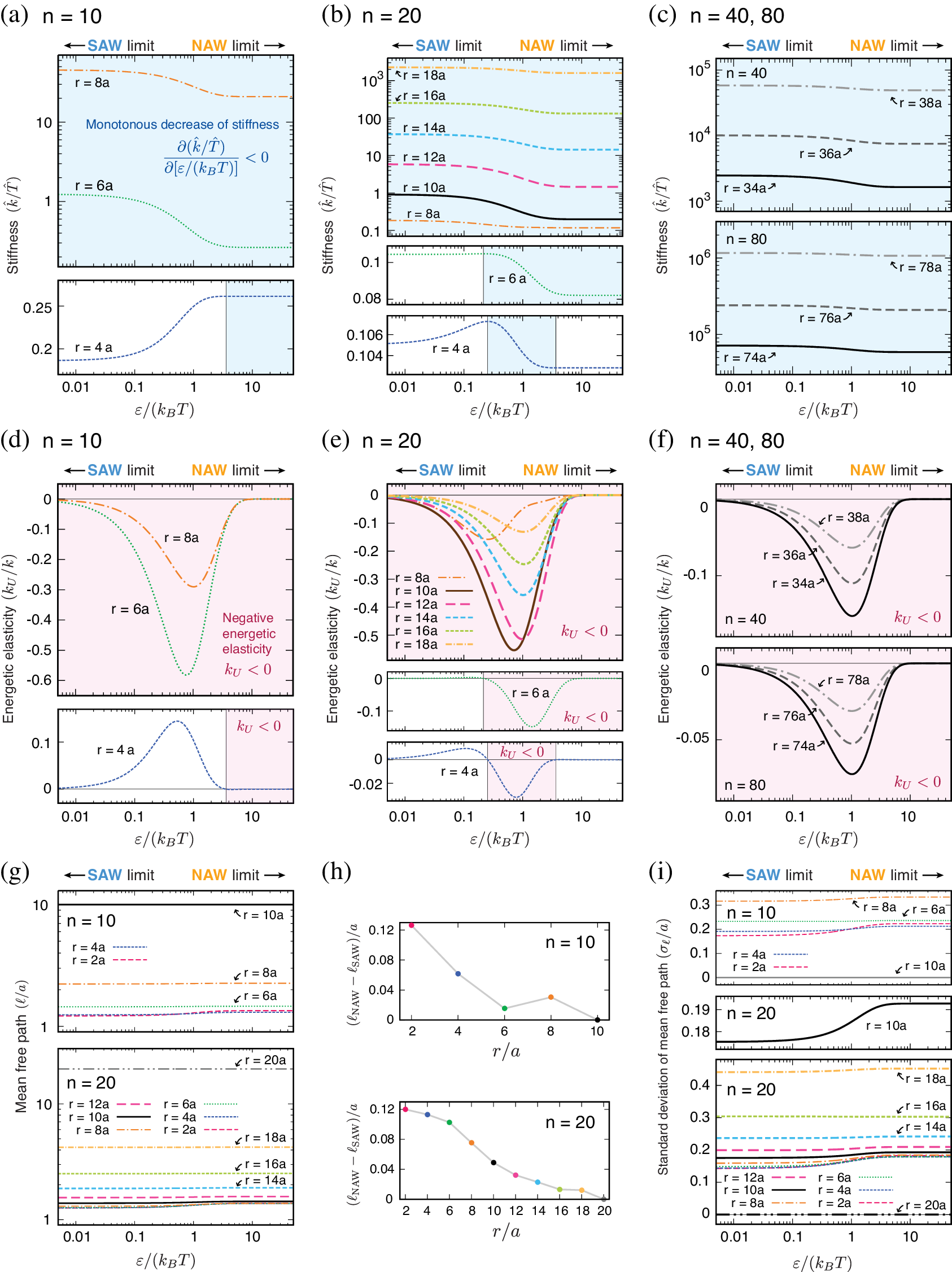}
\caption{\label{fig:n_r_dependencies}
(a--f) Polymer--solvent interaction [$\varepsilon/(k_B T)$] dependences of (a,b,c) stiffness ($\hat{k}/\hat{T}$) and (d,e,f) negative energetic elasticity $k_U/k$ for $\varepsilon>0$ in the logarithmic scale.
We show (a,d) $n=10$, (b,e) $n=20$, and (c,f) $n=40$ and $80$.
Curves for all possible $r$ are shown in (a), (b), (d), and (e), whereas curves for $r=(n-2)a$, $(n-4)a$, and $(n-6)a$ are shown in (c) and (f). 
The light-blue regions in (a), (b), and (c) exhibit the monotonically decreasing regions of $\hat{k}/\hat{T}$, i.e., $\partial (\hat{k}/\hat{T})/\partial [\varepsilon/(k_B T)]<0$. 
The pink regions in (d), (e), and (f) display negative energetic elasticity, i.e., $k_U<0$, which correspond to the light-blue regions in (a), (b), and (c), respectively, because of Eq.~(\ref{eq:decrease_in_k_over_T_and_k_U}). 
(g) Dependence of $\ell/a$ on 
$\varepsilon/(k_B T)$ for $(n,r)=(10,2a)$--$(10,10a)$ (top panel) and $(n,r)=(20,2a)$--$(20,20a)$ (bottom panel). 
(h) end-to-end distance ($r/a$) dependence of $(\ell_\mathrm{NAW} - \ell_\mathrm{SAW})/a$ for $(n,r)=(10,2a)$--$(10,10a)$ (top panel) and $(n,r)=(20,2a)$--$(20,20a)$ (bottom panel). 
(i) Dependence of $\sigma_\ell/a$ on $\varepsilon/(k_B T)$ for $(n,r)=(10,2a)$--$(10,10a)$ (top panel), for $(n,r)=(20,10a)$ (middle panel), and $(n,r)=(20,2a)$--$(20,20a)$ (bottom panel). 
Scales in the bottom panels of (a) and (b) and in all panels of (c) and (g) are double logarithmic scales.
}
\end{figure*}

Figures~\ref{fig:n_r_dependencies}(a,b) reveal the $\varepsilon/(k_B T)$ dependence of $\hat{k}/\hat{T}$ for all possible $r$ of $n=10$ and $20$ ($\varepsilon>0$).
The curves of $r\geq 6a$ for $n=10$ and $r\geq 8a$ for $n=20$ reveal a monotonic decrease in $\hat{k}/\hat{T}$ with $\varepsilon/(k_B T)$, i.e., 
\begin{eqnarray}
\frac{\partial (\hat{k}/\hat{T})}{\partial[\varepsilon/(k_B T)]}<0,
\label{eq:decrease_in_k_over_T}
\end{eqnarray}
whereas the curves for $(n,r)=(10,4a)$, $(20,4a)$, and $(20,6a)$ have increasing and decreasing regions of $\hat{k}/\hat{T}$.
Figure~\ref{fig:n_r_dependencies}(c) shows the $\varepsilon/(k_B T)$ dependence of $\hat{k}/\hat{T}$ for larger $n$ ($n=40$ and $80$) with $r=(n-2)a$, $(n-4)a$, and $r=(n-6)a$, which are exactly calculated using the polynomial functions $W_{n,m}\big((n-2)a\big)$, $W_{n,m}\big((n-4)a\big)$, and $W_{n,m}\big((n-6)a\big)$ given in Sec.~\ref{sec:W_r_eq_n_to_n_minus_8}. 
These curves monotonically decrease with $\varepsilon/(k_B T)$. 

The regions of Eq.~(\ref{eq:decrease_in_k_over_T}) correspond to those of $\hat{k}_U<0$ because
\begin{eqnarray}
\frac{\partial (\hat{k}/\hat{T})}{\partial[\varepsilon/(k_B T)]} &=& \frac{\partial (\hat{k}/\hat{T})}{\partial(1/\hat{T})} = \hat{k}-\hat{k}_S= \hat{k}_U,
\label{eq:decrease_in_k_over_T_and_k_U}
\end{eqnarray}
where we used $\varepsilon/(k_B T) \equiv 1/\hat{T}$, and Eqs.~(\ref{eq:k_S_def}) and (\ref{eq:f_0}).
Equation~(\ref{eq:decrease_in_k_over_T_and_k_U}) means that $\hat{k}/\hat{T}$ decreases when $\hat{k}_U$ is negative.

Figures~\ref{fig:n_r_dependencies}(d,e) show the $\varepsilon/(k_B T)$ dependence of $k_U/k$ for all possible $r$ of $n=10$ and $20$ ($\varepsilon>0$).
Notably, the sign of $k_U$ is the same as that of $k_U/k$ because $k>0$ for $\varepsilon/(k_B T)>0$ from our observation.
The curves of $r\geq 6a$ for $n=10$ and $r\geq 8a$ for $n=20$ exhibit $k_U<0$ for a positive finite $\varepsilon/(k_B T)$, i.e., $0<\varepsilon/(k_B T)<\infty$, whereas the curves for $(n,r)=(10,4a)$, $(20,4a)$, and $(20,6a)$ have the positive and negative regions of $k_U$.

Figure~\ref{fig:n_r_dependencies}(f) depicts the $\varepsilon/(k_B T)$ dependence of $k_U/k$ for larger $n$ ($n=40$ and $80$) with $r=(n-2)a$, $(n-4)a$, and $r=(n-6)a$, which are exactly calculated using the polynomial functions $W_{n,m}\big((r-2)a\big)$, $W_{n,m}\big((r-4)a\big)$, and $W_{n,m}\big((r-6)a\big)$ given in Sec.~\ref{sec:W_r_eq_n_to_n_minus_8}. 
The sign of $k_U$ is always negative for a positive finite $\varepsilon/(k_B T)$ $[0<\varepsilon/(k_B T)<\infty]$ in each pair of $(n,r)$.

In Sec.~\ref{sec:negativeness_of_k_U}, we demonstrate that (i) $\hat{k}_U\big((n-2)a,\hat{T}\big)<0$ for $n\geq 7$ and positive finite $\hat{T}$ $(i.e., 0<\hat{T}<\infty)$, (ii) $\hat{k}_U\big((n-4)a,\hat{T}\big)<0$ for $n\geq 10$ and positive finite $\hat{T}$, and (iii) $\hat{k}_U\big((n-6)a,\hat{T}\big)<0$ for $n\geq 13$ and positive finite $\hat{T}$. 
Thus, the lattice polymer chain model (interacting SAW) always exhibits negative energetic elasticity for $n\geq 13$ and positive finite $\hat{T}$, at least, around the fully stretched conditions [i.e., $r=(n-2)a$, $(n-4)a$, and $(n-6)a$]. 

Figure~\ref{fig:n_r_dependencies}(g) shows the dependencies of $\ell$ on $\varepsilon/(k_B T)$ of all possible $r$ for $n=10$ and $20$. 
For $(n,r)=(10,10a)$ and $(20,20a)$, both $\ell$ remain at $10a$ and $20a$, respectively, independent of $\varepsilon/(k_B T)$, because the fully stretched $\omega$ is only allowed. 
For $(n,r)=(10,2a)$--$(10,8a)$ and $(20,2a)$--$(20,18a)$, $\ell$ monotonically increases with $\varepsilon/(k_B T)$ and converges to finite values in the SAW and NAW limits. 
As indicated in Fig.~\ref{fig:n_r_dependencies}(h), the differences of $\ell$ between the NAW and SAW limits ($\ell_\mathrm{NAW}-\ell_\mathrm{SAW}$) are positive, except for the fully stretched conditions, i.e., $(n,r)=(10,10a)$ and $(20,20a)$, for which the values are zero. 

When $\hat{k}_U=\frac{\partial (\hat{k}/\hat{T})}{\partial[\varepsilon/(k_B T)]}<0$  [Eq.~(\ref{eq:decrease_in_k_over_T_and_k_U})], $\ell$ and $\hat{k}/\hat{T}$ are negatively correlated because $\ell$ monotonically increases with $\varepsilon/(k_B T)$ and $\hat{k}/\hat{T}$ monotonically decreases with $\varepsilon/(k_B T)$.

The variance and standard deviation of the mean free path $\ell$ are defined by
\begin{eqnarray} %
V_\ell &\equiv& \frac{1}{Z}
\sum_{m=0}^{m_\mathrm{ub}} \sum_{b=0}^{n-1}\left(\frac{na}{b+1}\right)^2W_{n,m,b}(r)\,e^{- \varepsilon m/(k_B T)}
- \ell^2, \notag\\
\sigma_\ell &\equiv& V_\ell^{\frac{1}{2}}, \label{eq:sigma_l}
\end{eqnarray}
where $\ell$ is given by Eq.~(\MainEql{}) in the main text. 
The middle panel of Fig.~\ref{fig:n_r_dependencies}(i) shows the $\varepsilon/(k_BT)$ dependence of $\sigma_\ell$ for $(n,r)=(20,10a)$ and $\varepsilon>0$. 
$\sigma_\ell$ increases with $\varepsilon/(k_BT)$, ranging from $0.175a$ to $0.193a$. 
Because $\ell_\mathrm{NAW}-\ell_\mathrm{SAW}\simeq 0.049a$ in Fig.~\ref{fig:n_r_dependencies}(h), the standard deviation of the mean free path $\ell$ is 3--4 times longer than $\ell_\mathrm{NAW}-\ell_\mathrm{SAW}$.
Thus, a larger effort is required to experimentally observe the negative correlation between $\ell$ and $\hat{k}/\hat{T}$ in a single polymer chain.
However, it might be observed in a polymer gel because it is composed of a large number of polymer chains.

The top and bottom panels of Fig.~\ref{fig:n_r_dependencies}(i) shows the $\varepsilon/(k_BT)$ dependence of $\sigma_\ell$ of all possible $r$ for $n=10$ and $20$, respectively.
$\sigma_\ell$ increases with $\varepsilon/(k_B T)$ for all $r$, except $(n,r)=(10,10a)$ and $(20,20a)$ because the fully stretched $\omega$ is only allowed in this condition.

\clearpage

\section{Relations between $\hat{k}_U^*$ and $\hat{T}_U^*$\label{sec:k_U_and_T_U}} 
This section provides relationships between $\hat{k}_U^*$ and $\hat{T}_U^*$.
The tangent line of $\hat{K}=\hat{k}(r, \hat{T})$ on the $\hat{T}$--$\hat{K}$ plane for a given $r$ at the reference temperature $\hat{T} = \hat{T}^*$ is
\begin{equation}
\label{eq:g_T}
\hat{K}=\hat{f}^*(\hat{T}) 
\equiv \hat{k}^* + \frac{\hat{k}_S^*}{\hat{T}^*}(\hat{T} - \hat{T}^*),
\end{equation}
where $\hat{k}^*\equiv \hat{k}(r,\hat{T}^*)$ is the stiffness at $\hat{T}=\hat{T}^*$ and 
\begin{equation}
\hat{k}_S^* \equiv \hat{T}^* \left.\frac{\partial \hat{k}(r,\hat{T})}{\partial \hat{T}}\right|_{\hat{T}=\hat{T}^*}
\label{eq:k_S_def}
\end{equation}
is the entropic contribution of $\hat{k}^*$.
The intercept of $\hat{K}=\hat{f}^*(\hat{T})$ on the vertical axis is
\begin{equation}
\label{eq:f_0}
\hat{f}^*(0) = \hat{k}^* - \hat{k}^*_S = \hat{k}^*_U \equiv \hat{k}_U(r,\hat{T}^*). 
\end{equation}

\begin{figure}[t!] 
\centering
\includegraphics[width=0.65\linewidth]{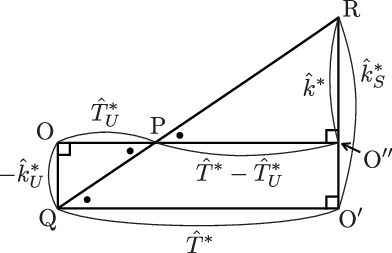}
\caption{\label{fig:similar_triangles}Similar triangles $\triangle \mathrm{OPQ}$, $\triangle \mathrm{O^\prime QR}$, and $\triangle \mathrm{O^{\prime\prime}PR}$ extracted from Fig.~\MainFigk{}~(b) for $\varepsilon>0$ and $-\hat{k}_U^*>0$. 
In the limit of $\hat{T}^*\to \infty$, triangles $\triangle \mathrm{O^\prime QR}$ and $\triangle \mathrm{O^{\prime\prime}PR}$ become infinitely large, whereas triangle $\triangle \mathrm{OPQ}$ converges to one with sides $\mathrm{OQ}=-\hat{k}_U^\infty$ and $\mathrm{OP}=\hat{T}_U^\infty$.}
\end{figure}

In the main text, we defined $\hat{T}_U^*$ as the intercept of the tangent line on the horizontal axis, i.e., $\hat{f}^*(\hat{T}_U^*)=0$. 
Using $\hat{f}^*(\hat{T}_U^*)=0$ and Eq.~(\ref{eq:g_T}), we can derive
\begin{equation} \label{eq:k_U_over_k}
\frac{-\hat{k}^*_U}{\hat{k}^*}=\frac{\hat{T}_U^*}{\hat{T}^*-\hat{T}_U^*},
\end{equation}
which relates $\hat{T}_U^*$ to $\hat{k}_U^*$. 
Additionally, we can derive
\begin{equation} \label{eq:hat_T_U_star}
\hat{T}_U^* = \frac{- \hat{k}_U^*}{\hat{k}_S^*}\hat{T}^* = \frac{-\hat{k}_U^*}{\left.\frac{\partial \hat{k}(r,\hat{T})}{\partial \hat{T}}\right|_{\hat{T}=\hat{T}^*}}.
\end{equation}
Equations~(\ref{eq:k_U_over_k}) and (\ref{eq:hat_T_U_star}) can be confirmed by the similar triangles in Fig.~\ref{fig:similar_triangles}.

\section{Analytic expression for entropic and energetic contributions of stiffness} 
This section provides the derivation of the analytic expressions for $\hat{k}_S$ and $\hat{k}_U$. 
Notably, we use $\hat{T}$ as a variable instead of $\hat{T}^*$. 
Using Eq.~(\MainEqStiffness{}), we derive
\begin{eqnarray}
\frac{\partial \hat{k}(r,\hat{T})}{\partial \hat{T}} &=&  \frac{1}{\hat{T}}\left[\hat{k} - \frac{2\varDelta Z}{Z^3\,\varDelta \hat{r}^2} \left(
Z\varDelta Z^\prime
-
\varDelta Z Z^\prime 
\right)\right. \notag\\
&& \quad \left. + \frac{1}{Z^2\,\varDelta \hat{r}^2}\left(
\varDelta^2 Z Z^\prime
-
Z \varDelta^2 Z^\prime
\right)\right],
\label{eq:del_k_over_del_T}
\end{eqnarray}
where
\begin{eqnarray*}
\varDelta \hat{r} &\equiv& \varDelta r/a = 2,\\
Z^\prime (r,\hat{T}) &\equiv& \frac{\partial Z(r,\hat{T})}{\partial(1/\hat{T})} = - \sum_{m=0}^{m_\mathrm{ub}} mW_{n,m}(r)\,e^{-m/\hat{T}}.
\end{eqnarray*}
We also defined
\begin{eqnarray*}
\varDelta Z(r,\hat{T}) &\equiv& \sum_{m=0}^{m_\mathrm{ub}} \varDelta W_{n,m}(r)\,e^{-m/\hat{T}},\\
\varDelta^2 Z(r,\hat{T}) &\equiv& \sum_{m=0}^{m_\mathrm{ub}} \varDelta^2 W_{n,m}(r)\,e^{- m/\hat{T}},\\
\varDelta Z^\prime(r,\hat{T}) &\equiv& \frac{\partial\varDelta Z(r,\hat{T})}{\partial(1/\hat{T})} = - \sum_{m=0}^{m_\mathrm{ub}} m\varDelta W_{n,m}(r)\,e^{- m/\hat{T}},\\
\varDelta^2 Z^\prime(r,\hat{T}) &\equiv& \frac{\partial\varDelta^2 Z(r,\hat{T})}{\partial(1/\hat{T})} = - \sum_{m=0}^{m_\mathrm{ub}} m\varDelta^2 W_{n,m}(r)\,e^{- m/\hat{T}},
\end{eqnarray*}
where $\varDelta W_{n,m}(r) \equiv \left[W_{n,m}(r+\varDelta r)- W_{n,m}(r-\varDelta r)\right]/2$, and
$\varDelta^2 W_{n,m}(r) \equiv W_{n,m}(r+\varDelta r) - 2W_{n,m}(r)+ W_{n,m}(r-\varDelta r)$.
From Eqs.~(\ref{eq:k_S_def}) and (\ref{eq:del_k_over_del_T}), we derive
\begin{eqnarray}
\hat{k}_S(r,\hat{T}) &=& \hat{k} - \left[
\frac{2\varDelta Z}{Z^3\,\varDelta \hat{r}^2} \left(
Z\varDelta Z^\prime
-
\varDelta Z Z^\prime 
\right)\right. \notag\\
&& \quad \left. + \frac{1}{Z^2\,\varDelta \hat{r}^2}\left(
\varDelta^2 Z Z^\prime
-
Z \varDelta^2 Z^\prime
\right)\right],
\label{eq:k_S}
\end{eqnarray}
Using Eqs.~(\ref{eq:f_0}) and (\ref{eq:k_S}), we derive
\begin{eqnarray}
\hat{k}_U(r,\hat{T}) &=& \frac{2\varDelta Z}{Z^3\,\varDelta \hat{r}^2} \left(
Z\varDelta Z^\prime
-
\varDelta Z Z^\prime 
\right) \notag\\
&& + \frac{1}{Z^2\,\varDelta \hat{r}^2}\left(
\varDelta^2 Z Z^\prime
-
Z \varDelta^2 Z^\prime
\right).\label{eq:k_U}
\end{eqnarray}

\section{Analytic expression for $\hat{T}_U^\infty(n,r)$ as a functional of $W_{n,m}(r)$} 
This section provides the derivation of the analytic expression for $\hat{T}_U^\infty(n,r)$ as a functional of $W_{n,m}(r)$ based on Eqs.~(\ref{eq:hat_T_U_star}) to calculate the exact rational numbers of $\hat{T}_U^\infty(n,r)$, which are listed in Tables~\ref{tab:hat_T_U_inf_n5to10}--\ref{tab:hat_T_U_inf_n11to20}. 
In the limit of $\hat{T}^*\to \infty$, both $\hat{k}_U^*$ and $\left.\frac{\partial \hat{k}(r,\hat{T})}{\partial \hat{T}}\right|_{\hat{T}=\hat{T}^*}$ in Eq.~(\ref{eq:hat_T_U_star}) converge. 
We define $\hat{k}_U^\infty(r)\equiv \lim_{\hat{T}^*\to \infty} \hat{k}_U^*$, as we have defined $\hat{T}_U^\infty(n,r)$ in the main text. 
From Eq.~(\ref{eq:k_U}), we obtain
\begin{eqnarray}
\hat{k}_U^\infty(r) &=& \frac{2\varDelta Z_\infty}{Z_\infty^3\,\varDelta \hat{r}^2} \left(
Z_\infty\varDelta Z_\infty^\prime
-
\varDelta Z_\infty Z_\infty^\prime 
\right) \notag\\
&+& \frac{1}{Z_\infty^2\,\varDelta \hat{r}^2}\left(
\varDelta^2 Z_\infty Z_\infty^\prime
-
Z_\infty \varDelta^2 Z_\infty^\prime
\right), \label{eq:k_U_inf}
\end{eqnarray}
where
\begin{eqnarray*}
Z_\infty &\equiv& \lim_{\hat{T}\to\infty} Z(r,\hat{T}) = \sum_{m=0}^{m_\mathrm{ub}} W_{n,m}(r),\\
Z_\infty^\prime &\equiv& \lim_{\hat{T}\to\infty}Z^\prime (r,\hat{T}) = - \sum_{m=0}^{m_\mathrm{ub}} mW_{n,m}(r),
\end{eqnarray*}

\begin{eqnarray*}
\varDelta Z_\infty &\equiv& \lim_{\hat{T}\to\infty}\varDelta Z(r,\hat{T}) =  \sum_{m=0}^{m_\mathrm{ub}} \varDelta W_{n,m}(r),\\
\varDelta^2 Z_\infty &\equiv& \lim_{\hat{T}\to\infty}\varDelta^2 Z(r,\hat{T}) =  \sum_{m=0}^{m_\mathrm{ub}} \varDelta^2 W_{n,m}(r),
\end{eqnarray*}

\begin{eqnarray*}
\varDelta Z_\infty^\prime &\equiv& \lim_{\hat{T}\to\infty}\varDelta Z^\prime(r,\hat{T}) =  - \sum_{m=0}^{m_\mathrm{ub}} m\varDelta W_{n,m}(r),\\
\varDelta^2 Z_\infty^\prime &\equiv& \lim_{\hat{T}\to\infty}\varDelta^2 Z^\prime(r,\hat{T}) =  - \sum_{m=0}^{m_\mathrm{ub}} m\varDelta^2 W_{n,m}(r).
\end{eqnarray*}
Because $W_{n,m}(r)$, $\varDelta W_{n,m}(r)$, $\varDelta^2 W_{n,m}(r)$, and $m$ are rational numbers, $\hat{k}_U^\infty(r)$ is a rational number for a given set of $n$ and $r$.
Moreover,
\begin{eqnarray} \label{eq:del_k_over_del_T_inf}
\lim_{\hat{T}\to\infty}
\frac{\partial \hat{k}(r,\hat{T})}{\partial \hat{T}}
&=& \lim_{\hat{T}\to\infty}\frac{\hat{k}_S(r,\hat{T})}{\hat{T}}\notag\\
&=& \left(\frac{\varDelta Z_\infty}{Z_\infty\,\varDelta \hat{r}}\right)^2 - \frac{\varDelta^2 Z_\infty}{Z_\infty\,\varDelta \hat{r}^2},
\end{eqnarray}
is a rational number for a given set of $n$ and $r$. 
Using Eqs.~(\ref{eq:hat_T_U_star}), (\ref{eq:k_U_inf}), and (\ref{eq:del_k_over_del_T_inf}), we obtain  $\hat{T}_U^\infty(n,r)$ as a functional of $W_{n,m}(r)$:
\begin{widetext}
\begin{equation}
\label{eq:exact_hat_T_U_inf}
\hat{T}_U^\infty(n,r) = \frac{
2\varDelta Z_\infty\left(
\varDelta Z_\infty Z_\infty^\prime - Z_\infty\varDelta Z_\infty^\prime\right) + Z_\infty\left(
Z_\infty \varDelta^2 Z_\infty^\prime
- \varDelta^2 Z_\infty Z_\infty^\prime
\right)
}
{
(\varDelta Z_\infty)^2 Z_\infty - Z_\infty^2 \varDelta^2 Z_\infty
},
\end{equation}
\end{widetext}
which is a rational number for a given set of $n$ and $r$.

\section{Discussion on data collapse} 
In this section, we discuss the physical interpretation of the data collapse shown in Fig.~\MainFigDataCollapse{}~(b) in the main text. 
We discuss the newly found scaling relation followed by the rescaling of $\varepsilon$ in the continuum limit ($n \to \infty$ and $a\to 0$).

\subsection{Scaling relation between $n$ and $s\equiv n-\hat{r}$}
We explain the scaling relation found in the data collapse shown in Fig.~\MainFigDataCollapse{}~(b) in the main text. 
For a lattice polymer chain with $n$ steps with the end-to-end distance $r$, at least $\hat{r}\equiv r/a$ steps are taken for the $+x$-direction and the rest steps are $s\equiv n-\hat{r}$. 
Some of $s$ steps are taken for the $\pm y$- and $\pm z$-directions to go away from and come back to the $x$-axis. 
In this sense, $s$ is the amount of slack in the chain. 
By using $s$, the horizontal axis of Fig.~\MainFigDataCollapse{}~(b) can be written as $(n-\hat{r})^\alpha/n=s^\alpha/n$. 

\begin{figure}[t!] 
\centering
\includegraphics[width=0.95\linewidth]{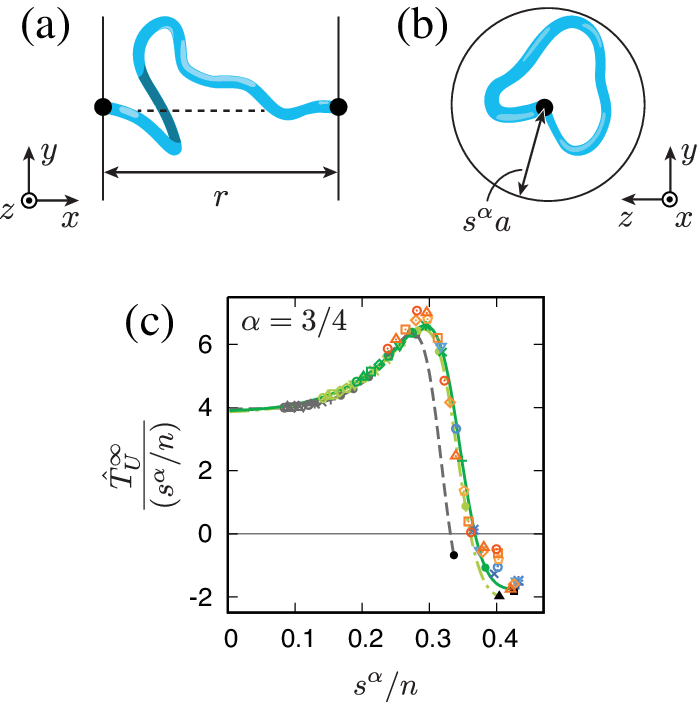}
\caption{\label{fig:TUinf_collapse_rescaling}
(a,b) Two views of a polymer chain on a three-dimensional space. 
(c) Rescaled plot of Fig.~\MainFigDataCollapse{}~(b) in the main text, to have a finite value of $T_U^\infty$ in the continuum limit ($n\to \infty$ and $a\to 0$).
}
\end{figure}

From the data collapse shown in Fig.~\MainFigDataCollapse{}~(b), we observed the scaling relation between two pairs of $n$ and $s$
\begin{equation}
\label{eq:scaling_relation}
s^\prime=\left(\frac{n^\prime}{n}\right)^{\frac{1}{\alpha}}s,
\end{equation}
where $\alpha=3/4$, which is the same as the universal critical exponent $\nu=3/4$ of the two-dimensional SAW~\cite{Nienhuis1982,Nienhuis1984}.
Equation~(\ref{eq:scaling_relation}) gives $s^\prime$ for $n^\prime$-step chain to have the same $\hat{T}_U^\infty$ as a chain with $n$ and $s$. 
Figures~\ref{fig:TUinf_collapse_rescaling}(a) and (b) illustrates a possible scenario to explain the origin of the critical exponent in a two-dimensional space ($\alpha=3/4$) as follows.
The on-axis constraint on the end-to-end vector reduces the effective dimensionality from three to two, and $s^\alpha a$, which contains the amount of chain slack $s$, gives typical length scale of the radius on the $yz$-plane.

The scaling relation in Eq.~(\ref{eq:scaling_relation}) can also be interpreted such that the sets of $(n,r)$ which have the same value of $s^\alpha/n$ have the same values of $\hat{T}_U^\infty$.
For example, the sets $(n,r)=(17,9a)$ and $(20,10a)$ that have almost the same values of $s^\alpha/n=(n-\hat{r})^\alpha/n\simeq 0.28$ give the almost the same values of $\hat{T}^\infty_U$ as
\begin{eqnarray*}
\hat{T}_U^\infty\big(17,9a\big)&=&\frac{3527424314659253062003}{1862575181539491665133}\simeq 1.89,\\
\hat{T}_U^\infty\big(20,10a\big)&=&\frac{2060069205296200981486403946}{1035875167962083022664474501}\simeq 1.99,
\end{eqnarray*}
where the rational numbers are taken from Table~\ref{tab:hat_T_U_inf_n11to20} in Sec.~\ref{sec:exact_values_of_TU_n5to20}.

\subsection{Rescaling of $\varepsilon$ in the continuum limit}
We discuss the rescaling of $\varepsilon$ proportional to $n$ to have the finite value of $T_U^\infty$ in the continuum limit ($n\to \infty$ and $a\to 0$). 
In this limit, $s^\alpha/n=(n-\hat{r})^\alpha/n\to 0$ and thus $\hat{T}_U^\infty=k_B T_U^\infty/\varepsilon \to 0$ [see Fig.~\MainFigDataCollapse{}~(b) in the main text].
However, assuming that the interaction energy $\varepsilon$ changes with $n$ as $\varepsilon\propto n$, the value of $T_U^\infty=\varepsilon\hat{T}_U^\infty/k_B$ can be rescued to be finite positive as shown in Fig.~\ref{fig:TUinf_collapse_rescaling}~(c).

Furthermore, we discuss the rescaling of $\varepsilon$ by using the rational functions given in Eqs.~(\ref{eq:hat_T_U_inf_rational_func_i2})--(\ref{eq:hat_T_U_inf_rational_func_i6}) in Sec.~\ref{sec:rational_functions}. 
The Taylor expansions of these functions by $1/n$ are given by
\begin{eqnarray}
\hat{T}_U^\infty\big(n,(n-2)a\big) &=& \frac{20}{3n} + O\left(\frac{1}{n^2}\right), \label{eq:TUinf_s2_expansion}\\
\hat{T}_U^\infty\big(n,(n-4)a\big) &=&
\frac{164}{15n} + O\left(\frac{1}{n^2}\right), \label{eq:TUinf_s4_expansion}\\
\hat{T}_U^\infty\big(n,(n-6)a\big) &=& \frac{524}{35n} +  O\left(\frac{1}{n^2}\right). \label{eq:TUinf_s6_expansion}
\end{eqnarray}
The leading terms of Eqs.~(\ref{eq:TUinf_s2_expansion})--(\ref{eq:TUinf_s6_expansion}) are in the first order of $1/n$. 
Thus, by multiplying the left-hand sides of Eqs.~(\ref{eq:TUinf_s2_expansion})--(\ref{eq:TUinf_s6_expansion}) by $n$ and taking the limit of $n\to \infty$,
we obtain the following finite positive values:
\begin{eqnarray}
\lim_{n\to\infty}
n\hat{T}_U^\infty\big(n,(n-2)a\big) &=& \frac{20}{3},\label{eq:nTUinf_s2_expansion}\\
\lim_{n\to\infty}
n\hat{T}_U^\infty\big(n,(n-4)a\big) &=&
\frac{164}{15},\label{eq:nTUinf_s4_expansion}\\
\lim_{n\to\infty}
n\hat{T}_U^\infty\big(n,(n-6)a\big) &=& \frac{524}{35}. \label{eq:nTUinf_s6_expansion}
\end{eqnarray}
These results suggest that $T_U^\infty=\varepsilon\hat{T}_U^\infty/k_B$ converges to a positive finite value if we assume $\varepsilon \propto n$.

By defining a rescaled interaction energy as $\varepsilon^\prime \equiv \varepsilon s^\alpha/n$ where $0<\varepsilon^\prime <\infty$, we obtain
\begin{eqnarray}
\frac{k_B}{\varepsilon^\prime}T_U^\infty\big(n,(n-2)a\big) &=& \frac{n}{2^\alpha}\hat{T}_U^\infty\big(n,(n-2)a\big)\notag\\
&=& \frac{20}{3 \cdot 2^\alpha} + O\left(\frac{1}{n}\right),\label{eq:nTUinf_s2_expansion2}\\
\frac{k_B}{\varepsilon^\prime}T_U^\infty\big(n,(n-4)a\big) &=&
\frac{n}{4^\alpha}\hat{T}_U^\infty\big(n,(n-4)a\big)\notag\\
&=&
\frac{164}{15 \cdot 4^\alpha} + O\left(\frac{1}{n}\right),\label{eq:nTUinf_s4_expansion2}\\
\frac{k_B}{\varepsilon^\prime}T_U^\infty\big(n,(n-6)a\big) &=&
\frac{n}{6^\alpha}\hat{T}_U^\infty\big(n,(n-6)a\big)\notag\\
&=& \frac{524}{35 \cdot 6^\alpha} + O\left(\frac{1}{n}\right).\label{eq:nTUinf_s6_expansion2}
\end{eqnarray}
In the limit of $n\to \infty$ with $\alpha=3/4$, the right-hand sides of Eqs.~(\ref{eq:nTUinf_s2_expansion2})--(\ref{eq:nTUinf_s6_expansion2}) converge to $20/(3\cdot 2^\alpha)\simeq 3.96$, $164/(15 \cdot 4^\alpha)\simeq 3.87$, and $524/(35 \cdot 6^\alpha)\simeq 3.91$, respectively. 
These values correspond to the $[\hat{T}_U^\infty/(s^\alpha/n)]$-intercepts of the curves in Fig.~\ref{fig:TUinf_collapse_rescaling}~(c). 
We expect that $k_B T_U^\infty\big(n,r\big)/\varepsilon^\prime$ converges to a single value ($\simeq 3.9$) independent of the value of $r$ in the continuum limit, which means that negative energetic elasticity exists even in the continuum limit. 
The rescaling of $\varepsilon$ is expected to be explained by using the renormalization group with simultaneously changing the length scale ($a$) and coupling constant ($\varepsilon$). 

\begin{figure}[b!] 
\centering
\includegraphics[width=0.5\linewidth]{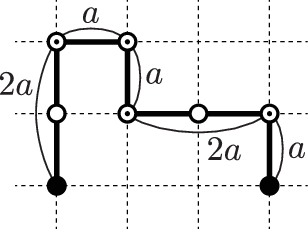}
\caption{\label{fig:mean_free_path}Example configuration $\omega$ for $(n,r)=(7,3a)$. 
The two endpoints are denoted by black circles ($\bullet$), and the bending points by circles with dots ($\odot$).
The number of bending points $b(\omega)$ is four.
}
\end{figure}

\section{Example of mean free path} 
Using the example configuration $\omega$ for $(n,r)=(7,3a)$ shown in Fig.~\ref{fig:mean_free_path}, we explain the thermal average of mean free path of the interacting SAW, $\ell(r,\hat{T})$, which is defined as Eq.~(\MainEql{}) in the main text.
In this configuration $\omega$, the number of bending points $b(\omega)$ is four. 
The number of straight paths between two bending points (between $\odot$ and $\odot$) and between an endpoint and a bending point (between $\odot$ and ${\bullet}$) is $b(\omega)+1=5$. 
Thus, the mean free path for $\omega$ is given by
\begin{equation}
\frac{2a+a+a+2a+a}{b(\omega)+1}=\frac{7}{5}a.
\label{eq:local_stiffness}
\end{equation}
Because the numerator of the left-hand side of Eq.~(\ref{eq:local_stiffness}) becomes $na$ for an arbitrary $\omega$, a mean free path can generally be rewritten as $na/(b(\omega)+1)$ for an arbitrary $n$, the thermal average of which yielded Eq.~(\MainEql{}) in the main text.

\section{SAW and NAW limits} 
The interacting SAW reduces to the SAW and NAW in the cases of $\left|\varepsilon\right|\ll k_{B}T$ and $\varepsilon\gg k_{B}T$, respectively.
This section explains the SAW ($\varepsilon \to 0$) and NAW ($\varepsilon \to \infty$) limits of the interacting SAW, as shown in Fig.~\MainFigLocalStiffness{} in the main text.

\subsection{Reduction to SAW ($\varepsilon \to 0$)}
In the limit of $\varepsilon \to 0$, Eq.~(\MainEqZ{}) in the main text becomes
\begin{equation}
Z(r,\infty) = \sum_{m=0}^{m_\mathrm{ub}} W_{n,m}(r) = W_n(r), \label{eq:Z_Tinf}
\end{equation}
because $\hat{T} \equiv k_{B}T/\varepsilon \to \infty$ and the Boltzmann factor becomes $e^{-m/\hat{T}}\to 1$ for each $\omega$. 
This means that the interacting SAW reduces to the SAW, which is defined as an ensemble of equally weighted $\omega$, in the limit of $\varepsilon \to 0$.
Thus, we call this limit as the SAW limit.

In the SAW limit, $\hat{k}(r,\hat{T})/\hat{T}$ converges to the following finite value:
\begin{equation}
\lim_{\hat{T}\to\infty}
\frac{\hat{k}(r,\hat{T})}{\hat{T}} = 
\left(\frac{\varDelta W_{n}(r)}{Z(r,\infty)\varDelta \hat{r}}  \right)^2 
- \frac{\varDelta^2 W_{n}(r)}{Z(r,\infty)\varDelta \hat{r}^2}. \label{eq:k_over_T_Tinf}
\end{equation}
For $(n,r)=(20,10a)$, we obtain
\begin{equation}
\lim_{\hat{T}\to\infty}
\frac{\hat{k}(10a,\hat{T})}{\hat{T}}
=\frac{6313217099066886233}{6892128429997467904}
\simeq 0.916,
\end{equation}
which is displayed in the top panel of Fig.~\MainFigLocalStiffness{} (dashed light-blue line labeled with ``SAW'') in the main text.

As discussed in Sec.~\ref{sec:k_U_and_T_U}, $\hat{k}_U(r,\hat{T})$ converges, in the SAW limit, to a finite rational number $\hat{k}_U^\infty(r,\hat{T})$ [Eq.~(\ref{eq:k_U_inf})], while Eq.~(\ref{eq:k_over_T_Tinf}) shows that $\hat{k}(r,\hat{T}) \to \infty$. 
Thus, we have
\begin{equation}
    \lim_{\hat{T}\to\infty}\frac{\hat{k}_U(r,\hat{T})}{\hat{k}(r,\hat{T})}=\lim_{\hat{T}\to\infty}\frac{k_U(r,\hat{T})}{k(r,\hat{T})}= 0,
\end{equation}
which means that the stiffness does not have an energetic contribution, and the elasticity of the lattice polymer chain is solely governed by entropy in the SAW limit.

\subsection{Reduction to NAW ($\varepsilon \to \infty$)}
In the limit of $\varepsilon \to \infty$, Eq.~(\MainEqZ{}) in the main text becomes
\begin{equation}
Z(r,0) = W_{n,0}(r), \label{eq:Z_T0}
\end{equation}
because $\hat{T} \equiv k_{B}T/\varepsilon \to 0$ and the Boltzmann factor becomes $e^{-m/\hat{T}}\to 0$, except for $m=0$. 
This means that the interacting SAW reduces to the NAW~\cite{TorrieWhittington1975,IshinabeChikahisa1986,NemirovskyDouglas1992}, which is defined as the SAW without first-neighbor contacts ($m=0$), in the limit of $\varepsilon \to \infty$, and thus we call this limit the NAW limit.

In the NAW limit, $\hat{k}(r,\hat{T})/\hat{T}$ converges to the following finite value:
\begin{eqnarray}
\lim_{\hat{T}\to 0}\frac{\hat{k}(r,\hat{T})}{\hat{T}} &=& 
\left(\frac{\varDelta W_{n,0}(r)}{Z(r,0)\varDelta \hat{r}}  \right)^2 
- \frac{\varDelta^2 W_{n,0}(r)}{Z(r,0)\varDelta \hat{r}^2}. \label{eq:k_over_T_T0}
\end{eqnarray}
For $(n,r)=(20,10a)$, we obtain
\begin{equation}
\lim_{\hat{T}\to 0} \frac{\hat{k}(10a,\hat{T})}{\hat{T}}=\frac{3389461192866292}{17118017290177281}\simeq 0.198,
\end{equation}
which is displayed in the top panel of Fig.~\MainFigLocalStiffness{} (dashed yellow line labeled with ``NAW'') in the main text.

From Eq.~(\ref{eq:k_U}), the
Taylor series of $\hat{k}_U(r,\hat{T})$ around $e^{-1/\hat{T}}=0$ ($\hat{T}=0$) is
\begin{eqnarray}
\hat{k}_U(r,\hat{T}) &=& 
\left(\frac{\varDelta^2 W_{n,1}}{W_{n,0}\varDelta\, \hat{r}^2} - \frac{W_{n,1}\varDelta^2 W_{n,0} + 2\varDelta W_{n,0}\varDelta W_{n,1}}{W_{n,0}^2 \varDelta\, \hat{r}^2} \right.\notag\\
&& \left.+ \frac{2W_{n,1}\varDelta W_{n,0}^2}{W_{n,0}^3\, \varDelta \hat{r}^2} \right)e^{-1/\hat{T}} + O(e^{-2/\hat{T}}). 
\label{eq:series_k_U}
\end{eqnarray}
Equation~(\ref{eq:series_k_U}) means that $\hat{k}_U(r,\hat{T})$ decays exponentially in the NAW limit.
By using Eqs.~(\ref{eq:k_over_T_T0}) and (\ref{eq:series_k_U}), we obtain
\begin{equation}
    \lim_{\hat{T}\to 0}\frac{\hat{k}_U(r,\hat{T})}{\hat{k}(r,\hat{T})}=
    \lim_{\hat{T}\to 0}\frac{k_U(r,\hat{T})}{k(r,\hat{T})} = 0,
\end{equation}
which means that the stiffness does not have an energetic contribution, and the elasticity of the lattice polymer chain is solely governed by entropy in the NAW limit.

\section{Tables of enumeration results} 
We provide complete lists of $W_n(r)$ and $W_{n,m}(r)$ for $n=1$--$20$ in Tables~\ref{tab:W_n_r_n1to19}--\ref{tab:W_n_r_m_n20}.
We also provide a list of $W_{n,m}\big((n-8)a\big)$ in Table~\ref{tab:W_n_r_m_n21to26}. 
Notably, each column in Tables~\ref{tab:W_n_r_n1to19}--\ref{tab:W_n_r_m_n21to26} has at least one value greater than $0$.
For $W_n(r)\geq 1$ (i.e., $\omega$ exists), $\hat{r}\equiv r/a$ is required to be odd in the case of odd $n$ (see Table~\ref{tab:W_n_r_n1to19}), and $\hat{r}$ is required to be even in the case of even $n$ (Table~\ref{tab:W_n_r_n2to20}).

\begin{widetext}

\input{list_W_n_r}

\end{widetext}

We list $W_{n,m}(r)$ in Tables~\ref{tab:W_n_r_m_n1to4}--\ref{tab:W_n_r_m_n20}. 
Because $W_{n,m}(r-\varDelta r)$ and $W_{n,m}(r+\varDelta r)$ for $\varDelta r\equiv 2a$ are required to calculate the stiffness $\hat{k}(r,\hat{T})$, the minimum set of $n$ and $r$ that provides the analytic expression for $\hat{k}(r,\hat{T})$ is $(n,r)=(5,3a)$, as shown in Table~\ref{tab:W_n_r_m_n5}.

\input{list_W_n_r_m_n1to16}

\begin{widetext}

\input{list_W_n_r_m_n17to20}

\input{list_W_n_r_m_n21to26}

\section{Polynomial functions $W_{n,m}\big((n-2)a\big)$, $W_{n,m}\big((n-4)a\big)$, $W_{n,m}\big((n-6)a\big)$, and $W_{n,m}\big((n-8)a\big)$\label{sec:W_r_eq_n_to_n_minus_8}} 

This section presents the analytic expressions for $W_{n,m}\big((n-2)a\big)$, $W_{n,m}\big((n-4)a\big)$, $W_{n,m}\big((n-6)a\big)$, and $W_{n,m}\big((n-8)a\big)$.

For $r=na$, only the fully stretched $\omega$ is allowed, and thus
\begin{eqnarray}
W_{n,0}(na) &=& 1 \qquad (n \ge 1), \label{eq:poly_W_i0_m0}\\
W_{n,m}(na) &=& 0 \qquad (n \ge 1,m \ge 1). \label{eq:poly_W_i0_m}
\end{eqnarray}

Assuming that $W_{n,m}\big((n-2)a\big)$ and $W_{n,m}\big((n-4)a\big)$ are polynomial functions of $n$, their exact forms are calculated using the values in Tables~\ref{tab:W_n_r_m_n1to4}--\ref{tab:W_n_r_m_n20} as
\begin{eqnarray}
W_{n,0}\big((n-2)a\big) &=& 2(n-2)(n-3) \qquad (n\ge 2), \label{eq:poly_W_i2_m0}\\
W_{n,1}\big((n-2)a\big) &=& 4(n-2) \qquad (n \ge 2), \label{eq:poly_W_i2_m1}\\
W_{n,m}\big((n-2)a\big) &=& 0 \qquad (n \ge 0,m \ge 2), \label{eq:poly_W_i2_m}\\
W_{n,0}\big((n-4)a\big) &=& \frac{1}{2}(3n^4 - 50n^3 + 353n^2 - 1266n + 1888) \qquad (n\ge 6), \label{eq:poly_W_i4_m0}\\
W_{n,1}\big((n-4)a\big) &=& 4(2n^3 - 28n^2 + 143n - 266) \qquad (n\ge 6), \label{eq:poly_W_i4_m1}\\
W_{n,2}\big((n-4)a\big) &=& 4(3n^2 - 26n + 67) \qquad (n\ge 6), \label{eq:poly_W_i4_m2}\\
W_{n,3}\big((n-4)a\big) &=& 4(n-6) \qquad (n\ge 6), \label{eq:poly_W_i4_m3}\\
W_{n,m}\big((n-4)a\big) &=& 0 \qquad (n\ge 1, m\ge 4). \label{eq:poly_W_i4_m}
\end{eqnarray}
We confirmed that all above equations produce the same values in Tables~\ref{tab:W_n_r_m_n1to4}--\ref{tab:W_n_r_m_n20} up to $n=20$.
We also confirmed them by additional enumeration results up to $n=87$.

The polynomial functions of $W_{n,m}\big((n-6)a\big)$ are also calculated using the values in Tables~\ref{tab:W_n_r_m_n10}--\ref{tab:W_n_r_m_n20} as
\begin{eqnarray}
W_{n,0}\big((n-6)a\big) &=& \frac{1}{9}(5n^6 - 183n^5 + 2999n^4 - 28461n^3 + 166160n^2 - 565248n + 865224) \qquad (n \ge 10), \label{eq:poly_W_i6_m0}\\
W_{n,1}\big((n-6)a\big) &=& 2(3n^5 - 98n^4 + 1367n^3 - 10236n^2 + 41244n - 71368)\qquad (n \ge 10), \label{eq:poly_W_i6_m1}\\
W_{n,2}\big((n-6)a\big) &=& 2(11n^4 - 289n^3 + 3018n^2 - 14966n + 29962)\qquad (n \ge 10), \label{eq:poly_W_i6_m2}\\
W_{n,3}\big((n-6)a\big) &=& \frac{8}{3}(13n^3 - 228n^2 + 1394n - 3027)\qquad (n \ge 10), \label{eq:poly_W_i6_m3}\\
W_{n,4}\big((n-6)a\big) &=& 2(11n^2 - 99n + 182)\qquad (n \ge 10), \label{eq:poly_W_i6_m4}\\
W_{n,5}\big((n-6)a\big) &=& 4(5n - 34)\qquad (n \ge 10), \label{eq:poly_W_i6_m5}\\
W_{n,m}\big((n-6)a\big) &=& 0 \qquad (n\ge 1 , m\ge 6). \label{eq:poly_W_i6_m}
\end{eqnarray}
We confirmed that all above equations yield the same values in Tables~\ref{tab:W_n_r_m_n10}--\ref{tab:W_n_r_m_n20} up to $n=20$.
We also confirmed them by extra enumeration results up to $n=35$. 

The polynomial functions of $W_{n,m}\big((n-8)a\big)$ are also calculated using the values in Tables~\ref{tab:W_n_r_m_n14}--\ref{tab:W_n_r_m_n21to26} as
\begin{eqnarray}
W_{n,0}\big((n-8)a\big) &=& \frac{1}{288}(35n^8 - 2260n^7 + 67334n^6 - 1214200n^5 + 14548595n^4 - 118874980n^3 \notag \\
&& \quad + 646859556n^2 - 2137772688n + 3268268928) \qquad (n \ge 14), \label{eq:poly_W_i8_m0}\\
W_{n,1}\big((n-8)a\big) &=& \frac{2}{9}(10n^7 - 596n^6 + 15931n^5 - 248384n^4 + 2448529n^3 - 15309446n^2 \notag\\
&& \quad + 56305152n - 93884472) \qquad (n \ge 14), \label{eq:poly_W_i8_m1}\\
W_{n,2}\big((n-8)a\big) &=& \frac{2}{3}(23n^6 - 1184n^5 + 26420n^4 - 328445n^3 + 2412356n^2 - 9985286n \notag\\
&& \quad + 18288954) \qquad (n \ge 14), \label{eq:poly_W_i8_m2}\\
W_{n,3}\big((n-8)a\big) &=& \frac{2}{3}(77n^5 - 3118n^4 + 51571n^3 - 437066n^2 + 1917102n - 3550464) \qquad (n \ge 14), \label{eq:poly_W_i8_m3}\\
W_{n,4}\big((n-8)a\big) &=& \frac{1}{3}(269n^4 - 7306n^3 + 66553n^2 - 191348n - 102132) \qquad (n \ge 14), \label{eq:poly_W_i8_m4}\\
W_{n,5}\big((n-8)a\big) &=& 8(17n^3 - 325n^2 + 1606n - 108) \qquad (n \ge 14), \label{eq:poly_W_i8_m5}\\
W_{n,6}\big((n-8)a\big) &=& 4(64n^2 - 1063n + 4321) \qquad (n \ge 14), \label{eq:poly_W_i8_m6}\\
W_{n,7}\big((n-8)a\big) &=& 4(73n - 698) \qquad (n \ge 14), \label{eq:poly_W_i8_m7}\\
W_{n,m}\big((n-8)a\big) &=& 0 \qquad (n \ge 1, m\ge 8). \label{eq:poly_W_i8_m}
\end{eqnarray}
We confirmed that all above equations produce the same values in Tables~\ref{tab:W_n_r_m_n14}--\ref{tab:W_n_r_m_n21to26} up to $n=26$.
We also confirmed them by additional enumeration results up to $n=31$.

\section{Negativeness of $\hat{k}_U$\label{sec:negativeness_of_k_U}} 

In this section, we demonstrate that $\hat{k}_U(r,\hat{T})<0$ for (A) $r=(n-2)a$ and $n \geq 7$, (B) $r=(n-4)a$ and $n\geq 10$, and (C) $r=(n-6)a$ and $n\geq 13$ in the case of positive finite $\hat{T}$, i.e.,  $0<\hat{T}<\infty$.

\subsection{Negativeness of $\hat{k}_U((n-2)a,\hat{T})$ for $n\geq 7$ and positive finite $\hat{T}$} 
This section shows $\hat{k}_U\big((n-2)a,\hat{T}\big) < 0$ for $n\geq 7$ and $0<\hat{T}<\infty$. 
Using Eq.~(\ref{eq:k_U}) and $W_{n,m}(r)$ for $r=na$, $(n-2)a$, and $(n-4)a$ given in Sec.~\ref{sec:W_r_eq_n_to_n_minus_8}, we derive
\begin{eqnarray}
\hat{k}_U\big((n-2)a, \hat{T}\big) &=& \frac{\sum_{m=1}^{7} u_m^{(2)}(n)\,e^{-m/\hat{T}}}{Z\big((n-2)a,\hat{T}\big)^3 \varDelta \hat{r}^2}, \label{eq:k_U_for_r_eq_n_minus_2}
\end{eqnarray}
where
\begin{eqnarray}
u_1^{(2)}(n) &=& -\frac{15}{2}n^9 + 269n^8 - 4353n^7 + 41736n^6 - \frac{521903}{2}n^5 + 1098783n^4 - 3089556n^3 + 5518880n^2\notag\\
&& \quad - 5584358n + 2400620, \label{eq:u_1_r_eq_n_minus_2}\\
u_2^{(2)}(n) &=& -76n^8 + 2420n^7 - 34308n^6 + 281984n^5 - 1462040n^4 + 4861908n^3 - 10030208n^2\notag\\
&& \quad + 11596568n - 5693008,\\
u_3^{(2)}(n) &=& -306n^7 + 8466n^6 - 102066n^5 + 692886n^4 - 2846496n^3 + 7027860n^2 - 9557376n\notag\\
&& \quad + 5452560,\\
u_4^{(2)}(n) &=& -620n^6 + 14664n^5 - 147476n^4 + 802152n^3 - 2470312n^2 + 4035648n - 2677216,\\
u_5^{(2)}(n) &=& -656n^5 + 13104n^4 - 106256n^3 + 433424n^2 - 876896n + 681472,\\
u_6^{(2)}(n) &=& -336n^4 + 5616n^3 - 34752n^2 + 93504n - 87552,\\
u_7^{(2)}(n) &=& -64(n - 2)(n - 6)^2. \label{eq:u_7_r_eq_n_minus_2}
\end{eqnarray}

The sign of $\hat{k}_U\big((n-2)a, \hat{T}\big)$ is determined by the signs of $u_m^{(2)}$ $(1\leq m \leq 7)$, because the denominator of Eq.~(\ref{eq:k_U_for_r_eq_n_minus_2}), i.e., $Z\big((n-2)a,\hat{T}\big)^3\varDelta \hat{r}^2$, is positive for $n \geq 7$. 
Eqs.~(\ref{eq:u_1_r_eq_n_minus_2})--(\ref{eq:u_7_r_eq_n_minus_2}) show that the signs of all the highest order terms of $u_m^{(2)}$ $(1\leq m \leq 7)$ are negative, which means that the numerator of Eq.~(\ref{eq:k_U_for_r_eq_n_minus_2}), i.e., $\sum_{m=1}^{7} u_m^{(2)}(n)\,e^{-m/\hat{T}}$, becomes negative for $n\geq n^{(2)}_\mathrm{max}$ and $0<\hat{T}<\infty$.
Here, $n^{(2)}_\mathrm{max}$ is the maximum real root of Eqs.~(\ref{eq:u_1_r_eq_n_minus_2})--(\ref{eq:u_7_r_eq_n_minus_2}). 
From Eq.~(\ref{eq:u_1_r_eq_n_minus_2}), we obtain $n^{(2)}_\mathrm{max}\simeq 6.059$. 
Thus, $\hat{k}_U\big((n-2)a,\hat{T}\big) < 0$ for $n\geq 7$ and $0<\hat{T}<\infty$.

\subsection{Negativeness of $\hat{k}_U((n-4)a,\hat{T})$ for $n\geq 10$ and positive finite $\hat{T}$} 
This section shows $\hat{k}_U\big((n-4)a,\hat{T}\big)<0$ for $n\geq 10$ and $0<\hat{T}<\infty$. 
Using Eq.~(\ref{eq:k_U}) and $W_{n,m}(r)$ for $r=(n-2)a$, $(n-4)a$, and $(n-6)a$ given in Sec.~\ref{sec:W_r_eq_n_to_n_minus_8}, we derive
\begin{equation}
\hat{k}_U\big((n-4)a, \hat{T}\big) = \frac{\sum_{m=1}^{13} u_m^{(4)}(n)\,e^{-m/\hat{T}}}{Z\big((n-4)a,\hat{T}\big)^3 \varDelta \hat{r}^2},
\label{eq:k_U_for_r_eq_n_minus_4}
\end{equation}
where
\begin{eqnarray}
u_{1}^{(4)} &=& \frac{1}{162} \big( -205n^{15} + 17363n^{14} - 698688n^{13} + 17730761n^{12} - 317419220n^{11} + 4245828881n^{10} \notag\\
&& \quad - 43821416090n^9 + 355164512947n^8 - 2277397470243n^7 + 11544526951456n^6 - 45855780255442n^5  \notag\\
&& \quad + 140117991125440n^4 - 318926398490256n^3 + 511004785243104n^2 - 516064526221536n  \notag\\
&& \quad + 247767166991232 \big), \label{eq:u_1_r_eq_n_minus_4}\\
u_{2}^{(4)} &=& \frac{1}{81} \big( -2292n^{14} + 182962n^{13} - 6898709n^{12} + 162941603n^{11} - 2693619010n^{10} + 32962849596n^9 \notag\\
&& \quad - 307814509095n^8 + 2227156036115n^7 - 12540489751504n^6 + 54667080390664n^5 - 181613304658406n^4 \notag\\
&& \quad + 446081581133124n^3 - 766436668515864n^2 + 825222400134768n - 419988439723968 \big),\\
u_{3}^{(4)} &=& \frac{1}{27} \big( -7519n^{13} + 558915n^{12} - 19501364n^{11} + 423083520n^{10} - 6368138721n^9 + 70212215259n^8 \notag\\
&& \quad - 583211737640n^7 + 3694063829592n^6 - 17837621361724n^5 + 64846633975794n^4 - 172529312064648n^3 \notag\\
&& \quad + 318184924389048n^2 - 364538996385120n + 195702377263488 \big),\\
u_{4}^{(4)} &=& \frac{1}{27} \big( -43084n^{12} + 2943944n^{11} - 93752620n^{10} + 1840731370n^9 - 24817313664n^8 + 241999362864n^7 \notag\\
&& \quad - 1749419009816n^6 + 9442153172062n^5 - 37743096655496n^4 + 108902191644560n^3 - 215118667552872n^2 \notag\\
&& \quad + 260853620100048n - 146470815055296 \big),\\
u_{5}^{(4)} &=& \frac{1}{27} \big( -160518n^{11} + 9936908n^{10} - 284229985n^9 + 4960184370n^8 - 58672327908n^7 + 493675756356n^6 \notag\\
&& \quad - 3012574289145n^5 + 13317744068638n^4 - 41734540576604n^3 + 88117321096560n^2 - 112491169475760n \notag\\
&& \quad + 65489934144960 \big),
\end{eqnarray}
\begin{eqnarray}
u_{6}^{(4)} &=& \frac{1}{9} \big( -138454n^{10} + 7648980n^9 - 193163868n^8 + 2936883228n^7 - 29757650838n^6 + 209746542876n^5 \notag\\
&& \quad - 1039937635256n^4 + 3573849415860n^3 - 8124045206544n^2 + 10986555333456n \notag\\
&& \quad - 6672578318208 \big),\\
u_{7}^{(4)} &=& \frac{1}{9} \big( -262602n^9 + 12793214n^8 - 281282902n^7 + 3664717478n^6 - 31169310400n^5 + 179313313244n^4 \notag\\
&& \quad - 696829903744n^3 + 1760929607712n^2 - 2620070309760n + 1743666655104 \big),\\
u_{8}^{(4)} &=& \frac{1}{9} \big( -380812n^8 + 16233984n^7 - 307961064n^6 + 3400535688n^5 - 23925651852n^4 + 109883795592n^3 \notag\\
&& \quad - 321827760272n^2 + 549918585792n - 420150423744 \big),\\
u_{9}^{(4)} &=& -47066n^7 + 1734106n^6 - 27881874n^5 + 254335006n^4 - 1424355060n^3 + 4904014744n^2 - 9624808720n \notag\\
&& \quad + 8326067904,\\
u_{10}^{(4)} &=& \frac{1}{3} \big( -116020n^6 + 3601912n^5 - 47217340n^4 + 336098232n^3 - 1374987136n^2 + 3073310816n \notag\\
&& \quad - 2941199808 \big),\\
u_{11}^{(4)} &=& \frac{1}{3} \big( -70024n^5 + 1820912n^4 - 19044392n^3 + 100891264n^2 - 272534112n + 300922368 \big),\\
u_{12}^{(4)} &=& -9840n^4 + 217872n^3 - 1797888n^2 + 6600768n - 9217536,\\
u_{13}^{(4)} &=& -64(n - 6)(5n - 34)^2. \label{eq:u_13_r_eq_n_minus_4}
\end{eqnarray}

The sign of $\hat{k}_U\big((n-4)a, \hat{T}\big)$ is determined by the signs of $u_m^{(4)}$ $(1\leq m \leq 13)$, because the denominator of Eq.~(\ref{eq:k_U_for_r_eq_n_minus_4}), i.e., $Z\big((n-4)a,\hat{T}\big)^3\varDelta \hat{r}^2$, is positive for $n \geq 10$. 
Eqs.~(\ref{eq:u_1_r_eq_n_minus_4})--(\ref{eq:u_13_r_eq_n_minus_4}) show that the signs of all the highest order terms of $u_m^{(4)}$ $(1\leq m \leq 10)$ are negative, which means that the numerator of Eq.~(\ref{eq:k_U_for_r_eq_n_minus_2}), i.e., $\sum_{m=1}^{13} u_m^{(4)}(n)\,e^{-m/\hat{T}}$, becomes negative for $n\geq n^{(4)}_\mathrm{max}$ and $0<\hat{T}<\infty$. 
Here, $n^{(4)}_\mathrm{max}$ is the maximum real root of Eqs.~(\ref{eq:u_1_r_eq_n_minus_4})--(\ref{eq:u_13_r_eq_n_minus_4}). 
From Eq.~(\ref{eq:u_1_r_eq_n_minus_4}), we obtain $n^{(4)}_\mathrm{max}\simeq 9.376$. 
Thus, $\hat{k}_U\big((n-4)a,\hat{T}\big) < 0$ for $n\geq 10$ and $0<\hat{T}<\infty$. 

\subsection{Negativeness of $\hat{k}_U((n-6)a,\hat{T})$ for $n\geq 13$ and positive finite $\hat{T}$} 
This section shows $\hat{k}_U\big((n-6)a,\hat{T}\big)<0$ for $n\geq 13$ and $0<\hat{T}<\infty$.
Using Eq.~(\ref{eq:k_U}) and $W_{n,m}(r)$ for $r=(n-4)a$, $r=(n-6)a$, and $(n-8)a$ given in Sec.~\ref{sec:W_r_eq_n_to_n_minus_8}, we derive
\begin{eqnarray}
\hat{k}_U\big((n-6)a, \hat{T}\big) &=& \frac{\sum_{m=1}^{19} u_m^{(6)}(n)\,e^{-m/\hat{T}}}{Z\big((n-6)a,\hat{T}\big)^3 \varDelta \hat{r}^2}, \label{eq:k_U_for_r_eq_n_minus_6}
\end{eqnarray}
where
\begin{eqnarray}
u_{1}^{(6)} &=& \frac{1}{746496} \big(-22925n^{21} + 3651350n^{20} - 280572525n^{19} + 13839826100n^{18} - 491835217542n^{17} \label{eq:u_1_r_eq_n_minus_6}\notag\\
&& \quad + 13394578705548n^{16} - 290258225825138n^{15} + 5128596084770864n^{14} - 75101845533459921n^{13} \notag\\
&& \quad + 921314498644022870n^{12} - 9530979229274591217n^{11} + 83414456661395135244n^{10} \notag\\
&& \quad - 617750432346814111772n^9 + 3860615533725157859096n^8 - 20237786401609541886192n^7 \notag\\
&& \quad + 88105767039420150604160n^6 - 313715610584983672619328n^5 + 892634669763201915829632n^4 \notag\\
&& \quad - 1957408417977989278040064n^3 + 3114100298779481453119488n^2 - 3206402212383458902425600n \notag\\
&& \quad + 1606922746557142444277760\big),
\end{eqnarray}
\begin{eqnarray}
u_{2}^{(6)} &=& \frac{1}{373248} \big(-474725n^{20} + 72477575n^{19} - 5323156610n^{18} + 250183929702n^{17} - 8441916150108n^{16} \notag\\
&& \quad + 217454214304098n^{15} - 4437896193992216n^{14} + 73495640587216832n^{13} - 1003320446829472673n^{12} \notag\\
&& \quad + 11404101758151396399n^{11} - 108541672144176036462n^{10} + 866834712312806494674n^9 \notag\\
&& \quad - 5800838459693683585190n^8 + 32368042499535569509328n^7 - 149224276510783406184848n^6 \notag\\
&& \quad + 560117593439229814433568n^5 - 1673174866998422091489120n^4 + 3837299622571001723063040n^3 \notag\\
&& \quad - 6362231683063664156659200n^2 + 6804281931945297919598592n \notag\\
&& \quad - 3531205492848809083109376\big),
\end{eqnarray}
\begin{eqnarray}
u_{3}^{(6)} &=& \frac{1}{20736} \big(-495875n^{19} + 72086460n^{18} - 5025247738n^{17} + 223394489829n^{16} \notag\\
&& \quad - 7102529220402n^{15} + 171653821513788n^{14} - 3271271052894104n^{13} + 50319169149151326n^{12} \notag\\
&& \quad - 634158942279803035n^{11} + 6607661982448768920n^{10} - 57176346809850992550n^9 \notag\\
&& \quad + 411032686072754768181n^8 - 2445957646395038035736n^7 + 11950350707040255519912n^6 \notag\\
&& \quad  - 47271964962020442274912n^5 + 148067212928307141442896n^4 - 354431490131617618928256n^3 \notag\\
&& \quad  + 610754838651846303941376n^2 - 676280903480102933084160n + 362159556514847639764992 \big),
\end{eqnarray}
\begin{eqnarray}
u_{4}^{(6)} &=& \frac{1}{20736} \big(-5620285n^{18} + 772132941n^{17} - 50682756822n^{16} + 2113006447828n^{15}\notag\\
&& \quad - 62726107326294n^{14} + 1408503972817966n^{13} - 24801704091456024n^{12} + 350274467388318700n^{11}\notag\\
&& \quad - 4023424777397486437n^{10} + 37880161507205000093n^9 - 293111471410540984602n^8\notag\\
&& \quad + 1860363913234590375512n^7 - 9617737330505538826208n^6 + 39973267634633782242512n^5\notag\\
&& \quad - 130711629673483090530080n^4 + 324730477367018637020160n^3 - 577635286547052572318208n^2 \notag\\
&& \quad + 657062120735043755065344n - 359962270693333721874432\big),
\end{eqnarray}
\begin{eqnarray}
u_{5}^{(6)} &=& \frac{1}{41472} \big(-86374239n^{17} + 11118030894n^{16} - 680751301052n^{15} + 26345637587448n^{14} \notag\\
&& \quad - 722078681087354n^{13} + 14879005868088436n^{12} - 238754668762854404n^{11} + 3048069473546590952n^{10} \notag\\
&& \quad - 31348779742809829039n^9 + 261261071604966089454n^8 - 1764566880882885053896n^7\notag\\
&& \quad + 9604722295689978919088n^6 - 41621561034169404074224n^5 + 140604985480011158758624n^4\notag\\
&& \quad - 357666125384940531806208n^3 + 645773077991961321868800n^2 - 739228426992453425504256n\notag\\
&& \quad + 404186906092371006947328\big),
\end{eqnarray}
\begin{eqnarray}
u_{6}^{(6)} &=& \frac{1}{216} \big(-2519109n^{16} + 301054954n^{15} - 17019940109n^{14} + 604426476467n^{13}\notag\\
&& \quad - 15095427145153n^{12} + 281172547605099n^{11} - 4040360987784411n^{10} + 45680526525048917n^9\notag\\
&& \quad - 410503399138529674n^8 + 2939888305754964247n^7 - 16706993064096336872n^6\notag\\
&& \quad + 74441420566390517484n^5 - 254437714984828683168n^4 + 643073952504833286432n^3\notag\\
&& \quad - 1128767985081512244864n^2 + 1221712905468316713984n - 608149305930026566656\big),
\end{eqnarray}
\begin{eqnarray}
u_{7}^{(6)} &=& \frac{1}{648} \big(-32756323n^{15} + 3607145648n^{14} - 186648376466n^{13} + 6020529786436n^{12} \notag\\
&& \quad - 135370208237880n^{11} + 2246572148940090n^{10} - 28405848149145284n^9 + 278273548599611260n^8 \notag\\
&& \quad - 2124940856914580017n^7 + 12606389734008460634n^6 - 57328372730768167542n^5\notag\\
&& \quad + 194569629878650336716n^4 - 469855782171501542952n^3 + 739277388220218599088n^2\notag\\
&& \quad - 628430759332455883968n + 160520919782147767296\big),
\end{eqnarray}
\begin{eqnarray}
u_{8}^{(6)} &=& \frac{1}{648} \big(-116440367n^{14} + 11779763789n^{13} - 556037454843n^{12} + 16231347437467n^{11} \notag\\
&& \quad - 327255408288211n^{10} + 4817510908935999n^9 - 53328847399804225n^8 + 449976144865454749n^7 \notag\\
&& \quad - 2897724989746956462n^6 + 14088956084126042156n^5 - 50397048291189084356n^4 \notag\\
&& \quad + 126197591177892469200n^3 - 200757496958428129488n^2 + 161745657184397195712n \notag\\
&& \quad - 23858838506519788032\big),
\end{eqnarray}
\begin{eqnarray}
u_{9}^{(6)} &=& \frac{1}{72} \big(-39449002n^{13} + 3671864639n^{12} - 158499153965n^{11} + 4203519175384n^{10} \notag\\
&& \quad - 76458728456214n^9 + 1007702453590959n^8 - 9903811141683077n^7 + 73500501112531654n^6 \notag\\
&& \quad - 411864338108951494n^5 + 1720428809657422244n^4 - 5203310623522143240n^3 \notag\\
&& \quad + 10775222610142696272n^2 - 13658680985546304576n + 7963232487942859776\big),
\end{eqnarray}
\begin{eqnarray}
u_{10}^{(6)} &=& \frac{1}{108} \big(-156836388n^{12} + 13422268873n^{11} - 529491083274n^{10} + 12751566309959n^9 \notag\\
&& \quad - 209201009835930n^8 + 2468825217441915n^7 - 21547960824813174n^6 + 140593197746717825n^5 \notag\\
&& \quad - 683107973403606858n^4 + 2420733604571102548n^3 - 5967244349696783544n^2 + 9232367726463329376n \notag\\
&& \quad - 6809357484417268224\big),
\end{eqnarray}
\begin{eqnarray}
u_{11}^{(6)} &=& \frac{1}{72} \big(-237415031n^{11} + 18513703473n^{10} - 659368462582n^9 + 14185093823038n^8 \notag\\
&& \quad - 205315602729831n^7 + 2105539038681241n^6 - 15664778769980476n^5 + 84883702466085928n^4 \notag\\
&& \quad - 329809005388563632n^3 + 879667384846802256n^2 - 1458163614590797248n \notag\\
&& \quad + 1145237089233286656\big),
\end{eqnarray}
\begin{eqnarray}
u_{12}^{(6)} &=& \frac{1}{12} \big(-76017017n^{10} + 5325037344n^9 - 167894082086n^8 + 3140374045696n^7 - 38628157495073n^6 \notag\\
&& \quad + 326783899052848n^5 - 1926419437175376n^4 + 7811811918869872n^3 - 20825643072250992n^2 \notag\\
&& \quad + 32878321047905856n - 23272624228564224\big),
\end{eqnarray}
\begin{eqnarray}
u_{13}^{(6)} &=& \frac{1}{9} \big(-93866834n^9 + 5858628640n^8 - 161643126752n^7 + 2585639077156n^6 - 26383735512470n^5 \notag\\
&& \quad + 177480372403756n^4 - 781186773484440n^3 + 2134445192018064n^2 - 3163028481951648n \notag\\
&& \quad + 1737395413622208\big),
\end{eqnarray}
\begin{eqnarray}
u_{14}^{(6)} &=& \frac{1}{9} \big(-135337400n^8 + 7556760704n^7 - 183598113112n^6 + 2536829481800n^5 - 21820249213168n^4 \notag\\
&& \quad + 119680667774120n^3 - 408247985819312n^2 + 787623607261920n - 648632386240704\big),
\end{eqnarray}
\begin{eqnarray}
u_{15}^{(6)} &=& -18486076n^7 + 922340348n^6 - 19684120052n^5 + 233814892852n^4 - 1678620044832n^3 + 7335440548704n^2 \notag\\
&& \quad - 18210926298720n + 19960569475392,
\end{eqnarray}
\begin{eqnarray}
u_{16}^{(6)} &=& -17558536n^6 + 762236576n^5 - 13720599480n^4 + 131631126624n^3 - 714423668032n^2 + 2098590900128n \notag\\
&& \quad - 2635200769344,
\end{eqnarray}
\begin{eqnarray}
u_{17}^{(6)} &=& \frac{1}{3} \big(-34417792n^5 + 1247212336n^4 - 17748700224n^3 + 123907713968n^2 - 424390789856n \notag\\
&& \quad + 570898285824 \big),\\
u_{18}^{(6)} &=& -5056272n^4 + 153070320n^3 - 1694712576n^2 + 8104545984n - 14100627456,\\
u_{19}^{(6)} &=& -(320n - 2176)(73n - 698)^2. \label{eq:u_19_r_eq_n_minus_6}
\end{eqnarray}

The sign of $\hat{k}_U\big((n-6)a, \hat{T}\big)$ is determined by the signs of $u_m^{(6)}$ $(1\leq m \leq 19)$, because the denominator of Eq.~(\ref{eq:k_U_for_r_eq_n_minus_6}), i.e., $Z\big((n-6)a,\hat{T}\big)^3\varDelta \hat{r}^2$, is positive for $n \geq 1$. 
Eqs.~(\ref{eq:u_1_r_eq_n_minus_6})--(\ref{eq:u_19_r_eq_n_minus_6}) show that the signs of all the highest order terms of $u_m^{(6)}$ $(1\leq m \leq 19)$ are negative, which means that the numerator of Eq.~(\ref{eq:k_U_for_r_eq_n_minus_2}), i.e., $\sum_{m=1}^{19} u_m^{(6)}(n)\,e^{-m/\hat{T}}$, becomes negative for $n\geq n^{(6)}_\mathrm{max}$.  Here, $n^{(6)}_\mathrm{max}$ is the maximum real root of Eqs.~(\ref{eq:u_1_r_eq_n_minus_6})--(\ref{eq:u_19_r_eq_n_minus_6}). 
From Eq.~(\ref{eq:u_1_r_eq_n_minus_6}), we obtain $n^{(6)}_\mathrm{max}\simeq 12.185$. 
Thus, $\hat{k}_U\big((n-6)a,\hat{T}\big) < 0$ for $n\geq 13$ and $0<\hat{T}<\infty$.

\section{Exact values of $\hat{T}_U^\infty(n,r)$ for $n=5$--$20$\label{sec:exact_values_of_TU_n5to20}} 

Tables~\ref{tab:hat_T_U_inf_n5to10}--\ref{tab:hat_T_U_inf_n11to20} list the exact rational numbers of $\hat{T}_U^\infty(n,r)$ calculated using Eq.~(\ref{eq:exact_hat_T_U_inf}) with Tables~\ref{tab:W_n_r_m_n1to4}--\ref{tab:W_n_r_m_n20}.

\input{list_hat_T_U_inf}

\section{Rational functions $\hat{T}_U^\infty\big(n,(n-2)a\big)$, $\hat{T}_U^\infty\big(n,(n-4)a\big)$, and $\hat{T}_U^\infty\big(n,(n-6)a\big)$\label{sec:rational_functions}} 

This section shows the analytic expressions for $\hat{T}_U^\infty\big(n,(n-2)a\big)$, $\hat{T}_U^\infty\big(n,(n-4)a\big)$, and $\hat{T}_U^\infty\big(n,(n-6)a\big)$. 
Using $W_{n,m}(na)$, $W_{n,m}\big((n-2)a\big)$, $W_{n,m}\big((n-4)a\big)$, $W_{n,m}\big((n-6)a\big)$, and $W_{n,m}\big((n-8)a\big)$ given in Sec.~\ref{sec:W_r_eq_n_to_n_minus_8}, we can derive the following rational functions:
\begin{eqnarray}
\hat{T}_U^\infty\big(n,(n-2)a\big) &=& \frac{4\,p_2(n)}{(n-1)q_2(n)}, \label{eq:hat_T_U_inf_rational_func_i2}\\ 
\hat{T}_U^\infty\big(n,(n-4)a\big) &=& \frac{4\, p_4(n)}{(n - 4)(3n^3 - 22n^2 + 65n - 62)\,q_4(n)}, \label{eq:hat_T_U_inf_rational_func_i4}\\ 
\hat{T}_U^\infty\big(n,(n-6)a\big) &=& \frac{4\,p_6(n)}{(5n^6 - 129n^5 + 1433n^4 - 8745n^3 + 30962n^2 - 60390n + 49320)\,q_6(n)}, 
\label{eq:hat_T_U_inf_rational_func_i6} 
\end{eqnarray}
where
\begin{eqnarray*}
p_2(n) &=& 15n^8 - 356n^7 + 3766n^6 - 23016n^5 + 88019n^4 - 213804n^3 + 317784n^2 - 256008n + 81484,\\
q_2(n) &=& 9n^8 - 204n^7 + 2026n^6 - 11648n^5 + 42733n^4 - 102444n^3 + 156272n^2 - 137656n + 53028,\\
p_4(n) &=& 205n^{15} - 12779n^{14} + 377878n^{13} - 7028329n^{12} + 91843642n^{11} - 891705537n^{10} + 6632586110n^9\\
&& \quad - 38412484159n^8 + 174308221229n^7 - 618529465700n^6 + 1698600661640n^5 - 3535919882472n^4\\
&& \quad + 5382532241712n^3 - 5629298377824n^2 + 3590676428544n - 1040974789248,\\
q_4(n) &=& 25n^{12} - 1290n^{11} + 30701n^{10} - 447138n^9 + 4450497n^8 - 31949658n^7 + 169814699n^6\\
&& \quad- 673658250n^5 + 1979577382n^4 - 4199479272n^3 + 6095057184n^2 - 5417182800n + 2218551120,\\
p_6(n) &=& 22925n^{21} - 2701900n^{20} + 153468875n^{19} - 5586295180n^{18} + 146134227132n^{17} - 2919787308972n^{16}\\
&& + 46223429811614n^{15} - 593623263577288n^{14} + 6280555247726285n^{13} - 55286737479509908n^{12}\\
&& + 407269495940367735n^{11} - 2516577055829714988n^{10} + 13034676945473036810n^9 - 56377142495508236116n^8\\
&& + 202149359931692608736n^7 - 594051500342010659680n^6 + 1406389826610734976288n^5\\
&& - 2614817373934251716928n^4 + 3671888987236959830016n^3 - 3653905354412312687616n^2\\
&& + 2288827258144328171520n - 675038669997797867520,\\
q_6(n) &=& 1225n^{16} - 113400n^{15} + 4954420n^{14} - 135769920n^{13} + 2614686966n^{12} - 37554353328n^{11} + 416423467068n^{10}\\
&& - 3638603477568n^9 + 25331854157561n^8 - 141042593225496n^7 + 626123307241416n^6 - 2193065510167104n^5\\
&& + 5940674302476560n^4 - 12025282548916608n^3 + 17136895425944832n^2 - 15333010803855360n\\
&& + 6470585477775360.
\end{eqnarray*}

We confirmed that Eqs.~(\ref{eq:hat_T_U_inf_rational_func_i2}), (\ref{eq:hat_T_U_inf_rational_func_i4}), and (\ref{eq:hat_T_U_inf_rational_func_i6}) provide the correct rational numbers, as listed in Tables~\ref{tab:hat_T_U_inf_n5to10} and \ref{tab:hat_T_U_inf_n11to20}, for $5\leq n \leq 20$. 
We also confirmed that Eq.~(\ref{eq:hat_T_U_inf_rational_func_i2}) yields the correct rational numbers up to $n=80$, Eq.~(\ref{eq:hat_T_U_inf_rational_func_i4}) up to $n=35$, and Eq.~(\ref{eq:hat_T_U_inf_rational_func_i6}) up to $n=31$, based on additional exact enumeration results.

\end{widetext}

%% file: list_W_n_r.tex
\begin{table}[H]
\centering
\caption{\label{tab:W_n_r_n1to19}List of $W_n(r)$ for $n=1,3,\dots,19$. Values in column $W_n(a)$ are confirmed by Ref.~\cite{ButeraComi1999} and those in row $n=13$ are confirmed by Ref.~\cite{DombWilmers1965}.}
\begin{tabular}{@{}*{12}{r} @{}}
\hline
\hline
\headercell{\\$n$} & & \multicolumn{10}{c@{}}{$r$}\\
\cmidrule(l){3-12}
 & & $a$ & $3a$ & $5a$ & $7a$ & $9a$ & $11a$ & $13a$ & $15a$ & $17a$ & $19a$\\
\hline
$1$ & & $1$ & $0$ & $0$ & $0$ & $0$ & $0$ & $0$ & $0$ & $0$ & $0$\\
$3$ & & $4$ & $1$ & $0$ & $0$ & $0$ & $0$ & $0$ & $0$ & $0$ & $0$\\
$5$ & & $44$ & $24$ & $1$ & $0$ & $0$ & $0$ & $0$ & $0$ & $0$ & $0$\\
$7$ & & $552$ & $516$ & $60$ & $1$ & $0$ & $0$ & $0$ & $0$ & $0$ & $0$\\
$9$ & & $8040$ & $8936$ & $2320$ & $112$ & $1$ & $0$ & $0$ & $0$ & $0$ & $0$\\
$11$ & & $127016$ & $154752$ & $61624$ & $6944$ & $180$ & $1$ & $0$ & $0$ & $0$ & $0$\\
$13$ & & $2112320$ & $2729100$ & $1408808$ & $272136$ & $16452$ & $264$ & $1$ & $0$ & $0$ & $0$\\
$15$ & & $36484128$ & $49090288$ & $29916784$ & $8362224$ & $903880$ & $33484$ & $364$ & $1$ & $0$ & $0$\\
$17$ & & $648529392$ & $898975960$ & $614123820$ & $221870344$ & $36713772$ & $2468712$ & $61256$ & $480$ & $1$ & $0$\\
$19$ & & $11790401800$ & $16720159740$ & $12409673296$ & $5391815300$ & $1222907256$ & $129356500$ & $5851096$ & $103560$ & $612$ & $1$\\
\hline
\hline
\end{tabular}
\end{table}

\begin{table}[H]
\centering
\caption{\label{tab:W_n_r_n2to20}List of $W_n(r)$ for $n=2,4,\dots,20$.}
\begin{tabular}{@{}*{12}{r} @{}}
\hline
\hline
\headercell{\\$n$} & & \multicolumn{10}{c@{}}{$r$}\\
\cmidrule(l){3-12}
 & & $2a$ & $4a$ & $6a$ & $8a$ & $10a$ & $12a$ & $14a$ & $16a$ & $18a$ & $20a$\\
\hline
$2$ & & $1$ & $0$ & $0$ & $0$ & $0$ & $0$ & $0$ & $0$ & $0$ & $0$\\
$4$ & & $12$ & $1$ & $0$ & $0$ & $0$ & $0$ & $0$ & $0$ & $0$ & $0$\\
$6$ & & $184$ & $40$ & $1$ & $0$ & $0$ & $0$ & $0$ & $0$ & $0$ & $0$\\
$8$ & & $2616$ & $1172$ & $84$ & $1$ & $0$ & $0$ & $0$ & $0$ & $0$ & $0$\\
$10$ & & $40744$ & $25180$ & $4164$ & $144$ & $1$ & $0$ & $0$ & $0$ & $0$ & $0$\\
$12$ & & $673040$ & $497812$ & $135232$ & $10936$ & $220$ & $1$ & $0$ & $0$ & $0$ & $0$\\
$14$ & & $11563632$ & $9562412$ & $3587520$ & $510516$ & $23840$ & $312$ & $1$ & $0$ & $0$ & $0$\\
$16$ & & $204708128$ & $182639672$ & $84977400$ & $18088032$ & $1524744$ & $45804$ & $420$ & $1$ & $0$ & $0$\\
$18$ & & $3709636344$ & $3496496500$ & $1892115152$ & $538054084$ & $70560108$ & $3858956$ & $80332$ & $544$ & $1$ & $0$\\
$20$ & & $68496828560$ & $67289840344$ & $40683432604$ & $14322531084$ & $2625286352$ & $227589504$ & $8638480$ & $131504$ & $684$ & $1$\\
\hline
\hline
\end{tabular}
\end{table}

%% file: list_W_n_r_m_n1to16.tex
\begin{table}[H]
\centering
\caption{\label{tab:W_n_r_m_n1to4}List of $W_{n,m}(r)$ for $n=1$--$4$.}
\begin{tabular}{rrrr}
\hline
\hline
$n$ & $r$ & $m$ & $W_{n,m}(r)$\\
\hline
1 & a & 0 & $1$\\
\hline
2 & 2a & 0 & $1$\\
\hline
3 & a & 1 & $4$\\
3 & 3a & 0 & $1$\\
\hline
4 & 2a & 0 & $4$\\
4 & 2a & 1 & $8$\\
4 & 4a & 0 & $1$\\
\hline
\hline
\end{tabular}
\end{table}

\begin{table}[H]
\centering
\caption{\label{tab:W_n_r_m_n5}List of $W_{n,m}(r)$ for $n=5$.}
\begin{tabular}{@{}*{5}{r} @{}}
\hline
\hline
\headercell{\\$m$} & \multicolumn{3}{c@{}}{$r$}\\
\cmidrule(l){3-5}
 & & $a$ & $3a$ & $5a$\\
\hline
$0$ & & $0$ & $12$ & $1$\\
$1$ & & $8$ & $12$ & $0$\\
$2$ & & $36$ & $0$ & $0$\\
\hline
\hline
\end{tabular}
\end{table}

\begin{table}[H]
\centering
\caption{\label{tab:W_n_r_m_n6}List of $W_{n,m}(r)$ for $n=6$.}
\begin{tabular}{@{}*{5}{r} @{}}
\hline
\hline
\headercell{\\$m$} & \multicolumn{3}{c@{}}{$r$}\\
\cmidrule(l){3-5}
 & & $2a$ & $4a$ & $6a$\\
\hline
$0$ & & $44$ & $24$ & $1$\\
$1$ & & $64$ & $16$ & $0$\\
$2$ & & $76$ & $0$ & $0$\\
\hline
\hline
\end{tabular}
\end{table}

\begin{table}[H]
\centering
\caption{\label{tab:W_n_r_m_n7}List of $W_{n,m}(r)$ for $n=7$.}
\begin{tabular}{@{}*{6}{r} @{}}
\hline
\hline
\headercell{\\$m$} & \multicolumn{4}{c@{}}{$r$}\\
\cmidrule(l){3-6}
 & & $a$ & $3a$ & $5a$ & $7a$\\
\hline
$0$ & & $0$ & $188$ & $40$ & $1$\\
$1$ & & $72$ & $196$ & $20$ & $0$\\
$2$ & & $128$ & $128$ & $0$ & $0$\\
$3$ & & $336$ & $4$ & $0$ & $0$\\
$4$ & & $0$ & $0$ & $0$ & $0$\\
$5$ & & $16$ & $0$ & $0$ & $0$\\
\hline
\hline
\end{tabular}
\end{table}

\begin{table}[H]
\centering
\caption{\label{tab:W_n_r_m_n8}List of $W_{n,m}(r)$ for $n=8$.}
\begin{tabular}{@{}*{6}{r} @{}}
\hline
\hline
\headercell{\\$m$} & \multicolumn{4}{c@{}}{$r$}\\
\cmidrule(l){3-6}
 & & $2a$ & $4a$ & $6a$ & $8a$\\
\hline
$0$ & & $364$ & $520$ & $60$ & $1$\\
$1$ & & $688$ & $440$ & $24$ & $0$\\
$2$ & & $724$ & $204$ & $0$ & $0$\\
$3$ & & $688$ & $8$ & $0$ & $0$\\
$4$ & & $120$ & $0$ & $0$ & $0$\\
$5$ & & $32$ & $0$ & $0$ & $0$\\
\hline
\hline
\end{tabular}
\end{table}

\begin{table}[H]
\centering
\caption{\label{tab:W_n_r_m_n9}List of $W_{n,m}(r)$ for $n=9$.}
\begin{tabular}{@{}*{7}{r} @{}}
\hline
\hline
\headercell{\\$m$} & \multicolumn{5}{c@{}}{$r$}\\
\cmidrule(l){3-7}
 & & $a$ & $3a$ & $5a$ & $7a$ & $9a$\\
\hline
$0$ & & $0$ & $1836$ & $1160$ & $84$ & $1$\\
$1$ & & $820$ & $2868$ & $844$ & $28$ & $0$\\
$2$ & & $1600$ & $2392$ & $304$ & $0$ & $0$\\
$3$ & & $1680$ & $1448$ & $12$ & $0$ & $0$\\
$4$ & & $3300$ & $344$ & $0$ & $0$ & $0$\\
$5$ & & $320$ & $48$ & $0$ & $0$ & $0$\\
$6$ & & $320$ & $0$ & $0$ & $0$ & $0$\\
\hline
\hline
\end{tabular}
\end{table}

\begin{table}[H]
\centering
\caption{\label{tab:W_n_r_m_n10}List of $W_{n,m}(r)$ for $n=10$.}
\begin{tabular}{@{}*{7}{r} @{}}
\hline
\hline
\headercell{\\$m$} & \multicolumn{5}{c@{}}{$r$}\\
\cmidrule(l){3-7}
 & & $2a$ & $4a$ & $6a$ & $8a$ & $10a$\\
\hline
$0$ & & $3976$ & $6416$ & $2264$ & $112$ & $1$\\
$1$ & & $7608$ & $8944$ & $1456$ & $32$ & $0$\\
$2$ & & $9408$ & $6204$ & $428$ & $0$ & $0$\\
$3$ & & $8960$ & $2968$ & $16$ & $0$ & $0$\\
$4$ & & $7376$ & $584$ & $0$ & $0$ & $0$\\
$5$ & & $2656$ & $64$ & $0$ & $0$ & $0$\\
$6$ & & $616$ & $0$ & $0$ & $0$ & $0$\\
$7$ & & $144$ & $0$ & $0$ & $0$ & $0$\\
\hline
\hline
\end{tabular}
\end{table}

\begin{table}[H]
\centering
\caption{\label{tab:W_n_r_m_n11}List of $W_{n,m}(r)$ for $n=11$.}
\begin{tabular}{@{}*{8}{r} @{}}
\hline
\hline
\headercell{\\$m$} & \multicolumn{6}{c@{}}{$r$}\\
\cmidrule(l){3-8}
 & & $a$ & $3a$ & $5a$ & $7a$ & $9a$ & $11a$\\
\hline
$0$ & & $0$ & $20160$ & $18344$ & $4024$ & $144$ & $1$\\
$1$ & & $9628$ & $35924$ & $23144$ & $2324$ & $36$ & $0$\\
$2$ & & $20784$ & $38940$ & $13812$ & $576$ & $0$ & $0$\\
$3$ & & $26112$ & $32044$ & $5392$ & $20$ & $0$ & $0$\\
$4$ & & $22432$ & $18564$ & $848$ & $0$ & $0$ & $0$\\
$5$ & & $33492$ & $7248$ & $84$ & $0$ & $0$ & $0$\\
$6$ & & $9648$ & $1440$ & $0$ & $0$ & $0$ & $0$\\
$7$ & & $4656$ & $432$ & $0$ & $0$ & $0$ & $0$\\
$8$ & & $0$ & $0$ & $0$ & $0$ & $0$ & $0$\\
$9$ & & $264$ & $0$ & $0$ & $0$ & $0$ & $0$\\
\hline
\hline
\end{tabular}
\end{table}

\begin{table}[H]
\centering
\caption{\label{tab:W_n_r_m_n12}List of $W_{n,m}(r)$ for $n=12$.}
\begin{tabular}{@{}*{8}{r} @{}}
\hline
\hline
\headercell{\\$m$} & \multicolumn{6}{c@{}}{$r$}\\
\cmidrule(l){3-8}
 & & $2a$ & $4a$ & $6a$ & $8a$ & $10a$ & $12a$\\
\hline
$0$ & & $47104$ & $77668$ & $45512$ & $6668$ & $180$ & $1$\\
$1$ & & $100400$ & $133304$ & $52240$ & $3496$ & $40$ & $0$\\
$2$ & & $127652$ & $131032$ & $27332$ & $748$ & $0$ & $0$\\
$3$ & & $133720$ & $92360$ & $8888$ & $24$ & $0$ & $0$\\
$4$ & & $116140$ & $45400$ & $1156$ & $0$ & $0$ & $0$\\
$5$ & & $87360$ & $14296$ & $104$ & $0$ & $0$ & $0$\\
$6$ & & $43288$ & $3032$ & $0$ & $0$ & $0$ & $0$\\
$7$ & & $12304$ & $720$ & $0$ & $0$ & $0$ & $0$\\
$8$ & & $4656$ & $0$ & $0$ & $0$ & $0$ & $0$\\
$9$ & & $416$ & $0$ & $0$ & $0$ & $0$ & $0$\\
\hline
\hline
\end{tabular}
\end{table}

\begin{table}[H]
\centering
\caption{\label{tab:W_n_r_m_n13}List of $W_{n,m}(r)$ for $n=13$.}
\begin{tabular}{@{}*{9}{r} @{}}
\hline
\hline
\headercell{\\$m$} & \multicolumn{7}{c@{}}{$r$}\\
\cmidrule(l){3-9}
 & & $a$ & $3a$ & $5a$ & $7a$ & $9a$ & $11a$ & $13a$\\
\hline
$0$ & & $0$ & $237688$ & $256604$ & $101232$ & $10460$ & $220$ & $1$\\
$1$ & & $120736$ & $486628$ & $418852$ & $106240$ & $5020$ & $44$ & $0$\\
$2$ & & $277200$ & $581980$ & $374616$ & $49368$ & $944$ & $0$ & $0$\\
$3$ & & $373548$ & $557096$ & $230512$ & $13664$ & $28$ & $0$ & $0$\\
$4$ & & $397656$ & $435816$ & $96336$ & $1508$ & $0$ & $0$ & $0$\\
$5$ & & $318696$ & $256008$ & $25636$ & $124$ & $0$ & $0$ & $0$\\
$6$ & & $363524$ & $121020$ & $5244$ & $0$ & $0$ & $0$ & $0$\\
$7$ & & $174104$ & $38280$ & $1008$ & $0$ & $0$ & $0$ & $0$\\
$8$ & & $66696$ & $13112$ & $0$ & $0$ & $0$ & $0$ & $0$\\
$9$ & & $12768$ & $1120$ & $0$ & $0$ & $0$ & $0$ & $0$\\
$10$ & & $7392$ & $352$ & $0$ & $0$ & $0$ & $0$ & $0$\\
\hline
\hline
\end{tabular}
\end{table}

\begin{table}[H]
\centering
\caption{\label{tab:W_n_r_m_n14}List of $W_{n,m}(r)$ for $n=14$.}
\begin{tabular}{@{}*{9}{r} @{}}
\hline
\hline
\headercell{\\$m$} & \multicolumn{7}{c@{}}{$r$}\\
\cmidrule(l){3-9}
 & & $2a$ & $4a$ & $6a$ & $8a$ & $10a$ & $12a$ & $14a$\\
\hline
$0$ & & $594808$ & $948628$ & $749232$ & $206400$ & $15700$ & $264$ & $1$\\
$1$ & & $1373736$ & $1932472$ & $1152400$ & $199088$ & $6944$ & $48$ & $0$\\
$2$ & & $1898572$ & $2227272$ & $940052$ & $83052$ & $1164$ & $0$ & $0$\\
$3$ & & $2057744$ & $1975688$ & $510808$ & $19928$ & $32$ & $0$ & $0$\\
$4$ & & $1929760$ & $1361568$ & $183208$ & $1904$ & $0$ & $0$ & $0$\\
$5$ & & $1569320$ & $712880$ & $42592$ & $144$ & $0$ & $0$ & $0$\\
$6$ & & $1123340$ & $282504$ & $7932$ & $0$ & $0$ & $0$ & $0$\\
$7$ & & $649680$ & $92840$ & $1296$ & $0$ & $0$ & $0$ & $0$\\
$8$ & & $244976$ & $24960$ & $0$ & $0$ & $0$ & $0$ & $0$\\
$9$ & & $99920$ & $2896$ & $0$ & $0$ & $0$ & $0$ & $0$\\
$10$ & & $16336$ & $704$ & $0$ & $0$ & $0$ & $0$ & $0$\\
$11$ & & $5440$ & $0$ & $0$ & $0$ & $0$ & $0$ & $0$\\
\hline
\hline
\end{tabular}
\end{table}

\begin{table}[H]
\centering
\caption{\label{tab:W_n_r_m_n15}List of $W_{n,m}(r)$ for $n=15$.}
\begin{tabular}{@{}*{10}{r} @{}}
\hline
\hline
\headercell{\\$m$} & \multicolumn{8}{c@{}}{$r$}\\
\cmidrule(l){3-10}
 & & $a$ & $3a$ & $5a$ & $7a$ & $9a$ & $11a$ & $13a$ & $15a$\\
\hline
$0$ & & $0$ & $3009072$ & $3372996$ & $1968216$ & $392056$ & $22724$ & $312$ & $1$\\
$1$ & & $1577216$ & $6777160$ & $6747032$ & $2843824$ & $349384$ & $9316$ & $52$ & $0$\\
$2$ & & $3876352$ & $8998524$ & $7424564$ & $2123176$ & $132044$ & $1408$ & $0$ & $0$\\
$3$ & & $5529184$ & $9275396$ & $6071444$ & $1027644$ & $27888$ & $36$ & $0$ & $0$\\
$4$ & & $6210048$ & $8100496$ & $3734916$ & $320816$ & $2344$ & $0$ & $0$ & $0$\\
$5$ & & $6013952$ & $6086036$ & $1725384$ & $65856$ & $164$ & $0$ & $0$ & $0$\\
$6$ & & $4691072$ & $3713508$ & $609528$ & $11104$ & $0$ & $0$ & $0$ & $0$\\
$7$ & & $4302848$ & $1946732$ & $182560$ & $1588$ & $0$ & $0$ & $0$ & $0$\\
$8$ & & $2608960$ & $805784$ & $41992$ & $0$ & $0$ & $0$ & $0$ & $0$\\
$9$ & & $1105632$ & $291500$ & $5312$ & $0$ & $0$ & $0$ & $0$ & $0$\\
$10$ & & $417920$ & $64816$ & $1056$ & $0$ & $0$ & $0$ & $0$ & $0$\\
$11$ & & $131584$ & $20880$ & $0$ & $0$ & $0$ & $0$ & $0$ & $0$\\
$12$ & & $13696$ & $0$ & $0$ & $0$ & $0$ & $0$ & $0$ & $0$\\
$13$ & & $5664$ & $384$ & $0$ & $0$ & $0$ & $0$ & $0$ & $0$\\
\hline
\hline
\end{tabular}
\end{table}

\begin{table}[H]
\centering
\caption{\label{tab:W_n_r_m_n16}List of $W_{n,m}(r)$ for $n=16$.}
\begin{tabular}{@{}*{10}{r} @{}}
\hline
\hline
\headercell{\\$m$} & \multicolumn{8}{c@{}}{$r$}\\
\cmidrule(l){3-10}
 & & $2a$ & $4a$ & $6a$ & $8a$ & $10a$ & $12a$ & $14a$ & $16a$\\
\hline
$0$ & & $7821504$ & $12212568$ & $10901768$ & $4724180$ & $702344$ & $31904$ & $364$ & $1$\\
$1$ & & $19504008$ & $27831016$ & $21100384$ & $6413072$ & $581104$ & $12184$ & $56$ & $0$\\
$2$ & & $28777856$ & $36497708$ & $21977636$ & $4399932$ & $200532$ & $1676$ & $0$ & $0$\\
$3$ & & $33310408$ & $36368520$ & $16533040$ & $1912128$ & $37752$ & $40$ & $0$ & $0$\\
$4$ & & $32617684$ & $29835856$ & $9138472$ & $525892$ & $2828$ & $0$ & $0$ & $0$\\
$5$ & & $28397288$ & $20304088$ & $3743224$ & $96160$ & $184$ & $0$ & $0$ & $0$\\
$6$ & & $22173972$ & $11379596$ & $1190508$ & $14788$ & $0$ & $0$ & $0$ & $0$\\
$7$ & & $15508488$ & $5231544$ & $317896$ & $1880$ & $0$ & $0$ & $0$ & $0$\\
$8$ & & $9539200$ & $2100120$ & $64984$ & $0$ & $0$ & $0$ & $0$ & $0$\\
$9$ & & $4399840$ & $667968$ & $8080$ & $0$ & $0$ & $0$ & $0$ & $0$\\
$10$ & & $1918048$ & $166928$ & $1408$ & $0$ & $0$ & $0$ & $0$ & $0$\\
$11$ & & $533344$ & $41056$ & $0$ & $0$ & $0$ & $0$ & $0$ & $0$\\
$12$ & & $191816$ & $1936$ & $0$ & $0$ & $0$ & $0$ & $0$ & $0$\\
$13$ & & $10288$ & $768$ & $0$ & $0$ & $0$ & $0$ & $0$ & $0$\\
$14$ & & $4384$ & $0$ & $0$ & $0$ & $0$ & $0$ & $0$ & $0$\\
\hline
\hline
\end{tabular}
\end{table}

%% file: list_W_n_r_m_n17to20.tex
\begin{table}[H]
\centering
\caption{\label{tab:W_n_r_m_n17}List of $W_{n,m}(r)$ for $n=17$.}
\begin{tabular}{@{}*{11}{r} @{}}
\hline
\hline
\headercell{\\$m$} & \multicolumn{9}{c@{}}{$r$}\\
\cmidrule(l){3-11}
 & & $a$ & $3a$ & $5a$ & $7a$ & $9a$ & $11a$ & $13a$ & $15a$ & $17a$\\
\hline
$0$ & & $0$ & $39764128$ & $44964768$ & $32228788$ & $10499244$ & $1197872$ & $43648$ & $420$ & $1$\\
$1$ & & $21154416$ & $97559200$ & $102972676$ & $59829356$ & $13411912$ & $924320$ & $15596$ & $60$ & $0$\\
$2$ & & $55713456$ & $140059308$ & $132363900$ & $58712968$ & $8491940$ & $293232$ & $1968$ & $0$ & $0$\\
$3$ & & $84656088$ & $156300500$ & $126562532$ & $40650300$ & $3338020$ & $49728$ & $44$ & $0$ & $0$\\
$4$ & & $100150296$ & $146333032$ & $96997244$ & $20279164$ & $817180$ & $3356$ & $0$ & $0$ & $0$\\
$5$ & & $100964340$ & $120285900$ & $60336716$ & $7426756$ & $134320$ & $204$ & $0$ & $0$ & $0$\\
$6$ & & $91093176$ & $88271332$ & $30616792$ & $2123436$ & $18984$ & $0$ & $0$ & $0$ & $0$\\
$7$ & & $70218564$ & $55777236$ & $12910496$ & $512960$ & $2172$ & $0$ & $0$ & $0$ & $0$\\
$8$ & & $56518488$ & $30954840$ & $4638412$ & $93656$ & $0$ & $0$ & $0$ & $0$ & $0$\\
$9$ & & $37248768$ & $14805436$ & $1364460$ & $11200$ & $0$ & $0$ & $0$ & $0$ & $0$\\
$10$ & & $18642576$ & $6103424$ & $320912$ & $1760$ & $0$ & $0$ & $0$ & $0$ & $0$\\
$11$ & & $8432208$ & $2032992$ & $68832$ & $0$ & $0$ & $0$ & $0$ & $0$ & $0$\\
$12$ & & $2676096$ & $611208$ & $4928$ & $0$ & $0$ & $0$ & $0$ & $0$ & $0$\\
$13$ & & $869184$ & $87584$ & $1152$ & $0$ & $0$ & $0$ & $0$ & $0$ & $0$\\
$14$ & & $185904$ & $29840$ & $0$ & $0$ & $0$ & $0$ & $0$ & $0$ & $0$\\
$15$ & & $0$ & $0$ & $0$ & $0$ & $0$ & $0$ & $0$ & $0$ & $0$\\
$16$ & & $5832$ & $0$ & $0$ & $0$ & $0$ & $0$ & $0$ & $0$ & $0$\\
\hline
\hline
\end{tabular}
\end{table}

\begin{table}[H]
\centering
\caption{\label{tab:W_n_r_m_n18}List of $W_{n,m}(r)$ for $n=18$.}
\begin{tabular}{@{}*{11}{r} @{}}
\hline
\hline
\headercell{\\$m$} & \multicolumn{9}{c@{}}{$r$}\\
\cmidrule(l){3-11}
 & & $2a$ & $4a$ & $6a$ & $8a$ & $10a$ & $12a$ & $14a$ & $16a$ & $18a$\\
\hline
$0$ & & $105557056$ & $163533284$ & $153739144$ & $87726404$ & $21848108$ & $1959472$ & $58400$ & $480$ & $1$\\
$1$ & & $283974456$ & $408610136$ & $348844984$ & $155529880$ & $26318384$ & $1415920$ & $19600$ & $64$ & $0$\\
$2$ & & $446465328$ & $587318928$ & $434990300$ & $143569636$ & $15442980$ & $415388$ & $2284$ & $0$ & $0$\\
$3$ & & $544990496$ & $644715240$ & $395913304$ & $91651520$ & $5527752$ & $64024$ & $48$ & $0$ & $0$\\
$4$ & & $564515160$ & $584576932$ & $282517004$ & $41459136$ & $1215576$ & $3928$ & $0$ & $0$ & $0$\\
$5$ & & $515513352$ & $456771088$ & $161006088$ & $13663600$ & $181152$ & $224$ & $0$ & $0$ & $0$\\
$6$ & & $427703048$ & $310538988$ & $74138476$ & $3528964$ & $23692$ & $0$ & $0$ & $0$ & $0$\\
$7$ & & $325579536$ & $183082896$ & $28590880$ & $780352$ & $2464$ & $0$ & $0$ & $0$ & $0$\\
$8$ & & $225133088$ & $92449368$ & $9235460$ & $127808$ & $0$ & $0$ & $0$ & $0$ & $0$\\
$9$ & & $141633072$ & $41790192$ & $2492792$ & $14672$ & $0$ & $0$ & $0$ & $0$ & $0$\\
$10$ & & $74301728$ & $15994960$ & $530816$ & $2112$ & $0$ & $0$ & $0$ & $0$ & $0$\\
$11$ & & $35271344$ & $5341952$ & $106448$ & $0$ & $0$ & $0$ & $0$ & $0$ & $0$\\
$12$ & & $13066448$ & $1427928$ & $7920$ & $0$ & $0$ & $0$ & $0$ & $0$ & $0$\\
$13$ & & $4740672$ & $284320$ & $1536$ & $0$ & $0$ & $0$ & $0$ & $0$ & $0$\\
$14$ & & $896808$ & $57424$ & $0$ & $0$ & $0$ & $0$ & $0$ & $0$ & $0$\\
$15$ & & $287392$ & $2864$ & $0$ & $0$ & $0$ & $0$ & $0$ & $0$ & $0$\\
$16$ & & $7360$ & $0$ & $0$ & $0$ & $0$ & $0$ & $0$ & $0$ & $0$\\
\hline
\hline
\end{tabular}
\end{table}

\begin{table}[H]
\centering
\caption{\label{tab:W_n_r_m_n19}List of $W_{n,m}(r)$ for $n=19$.}
\begin{tabular}{@{}*{12}{r} @{}}
\hline
\hline
\headercell{\\$m$} & \multicolumn{10}{c@{}}{$r$}\\
\cmidrule(l){3-12}
 & & $a$ & $3a$ & $5a$ & $7a$ & $9a$ & $11a$ & $13a$ & $15a$ & $17a$ & $19a$\\
\hline
$0$ & & $0$ & $539644712$ & $614989480$ & $489935816$ & $221566504$ & $42963092$ & $3092360$ & $76640$ & $544$ & $1$\\
$1$ & & $289146680$ & $1436384752$ & $1561813664$ & $1088610836$ & $374539656$ & $48915120$ & $2100328$ & $24244$ & $68$ & $0$\\
$2$ & & $816647760$ & $2213290760$ & $2238723264$ & $1306547904$ & $325266284$ & $26706432$ & $572772$ & $2624$ & $0$ & $0$\\
$3$ & & $1320107200$ & $2629974996$ & $2407605840$ & $1125835304$ & $191950044$ & $8758588$ & $80848$ & $52$ & $0$ & $0$\\
$4$ & & $1645486880$ & $2633684132$ & $2104249424$ & $746960872$ & $79112188$ & $1744128$ & $4544$ & $0$ & $0$ & $0$\\
$5$ & & $1740933380$ & $2313106024$ & $1555259428$ & $390423916$ & $23605412$ & $237472$ & $244$ & $0$ & $0$ & $0$\\
$6$ & & $1629116640$ & $1830756116$ & $984169648$ & $163847856$ & $5547364$ & $28912$ & $0$ & $0$ & $0$ & $0$\\
$7$ & & $1396670120$ & $1324646200$ & $535316180$ & $57575556$ & $1131392$ & $2756$ & $0$ & $0$ & $0$ & $0$\\
$8$ & & $1069934560$ & $855899664$ & $251985488$ & $16948372$ & $167448$ & $0$ & $0$ & $0$ & $0$ & $0$\\
$9$ & & $803670780$ & $495621016$ & $103628472$ & $4150852$ & $18500$ & $0$ & $0$ & $0$ & $0$ & $0$\\
$10$ & & $537244240$ & $258059204$ & $37078256$ & $812144$ & $2464$ & $0$ & $0$ & $0$ & $0$ & $0$\\
$11$ & & $301703960$ & $119185060$ & $11304984$ & $153040$ & $0$ & $0$ & $0$ & $0$ & $0$ & $0$\\
$12$ & & $150357520$ & $47592272$ & $2873456$ & $10912$ & $0$ & $0$ & $0$ & $0$ & $0$ & $0$\\
$13$ & & $59180360$ & $16375864$ & $572544$ & $1920$ & $0$ & $0$ & $0$ & $0$ & $0$ & $0$\\
$14$ & & $23607840$ & $4675960$ & $94336$ & $0$ & $0$ & $0$ & $0$ & $0$ & $0$ & $0$\\
$15$ & & $5160000$ & $1136720$ & $8832$ & $0$ & $0$ & $0$ & $0$ & $0$ & $0$ & $0$\\
$16$ & & $1112640$ & $96040$ & $0$ & $0$ & $0$ & $0$ & $0$ & $0$ & $0$ & $0$\\
$17$ & & $321240$ & $30248$ & $0$ & $0$ & $0$ & $0$ & $0$ & $0$ & $0$ & $0$\\
\hline
\hline
\end{tabular}
\end{table}

\begin{table}[H]
\centering
\caption{\label{tab:W_n_r_m_n20}List of $W_{n,m}(r)$ for $n=20$.}
\begin{tabular}{@{}*{12}{r} @{}}
\hline
\hline
\headercell{\\$m$} & \multicolumn{10}{c@{}}{$r$}\\
\cmidrule(l){3-12}
 & & $2a$ & $4a$ & $6a$ & $8a$ & $10a$ & $12a$ & $14a$ & $16a$ & $18a$ & $20a$\\
\hline
$0$ & & $1452657120$ & $2245560848$ & $2172427044$ & $1457206084$ & $523343364$ & $80444436$ & $4730696$ & $98884$ & $612$ & $1$\\
$1$ & & $4204845176$ & $6100416256$ & $5550780112$ & $3145134240$ & $843385216$ & $86762704$ & $3030224$ & $29576$ & $72$ & $0$\\
$2$ & & $7031181480$ & $9473908684$ & $7864625864$ & $3615527208$ & $689830188$ & $44243756$ & $771684$ & $2988$ & $0$ & $0$\\
$3$ & & $9039457896$ & $11195485032$ & $8224333568$ & $2940899072$ & $377456344$ & $13368784$ & $100408$ & $56$ & $0$ & $0$\\
$4$ & & $9813883384$ & $11012590952$ & $6883346012$ & $1813973596$ & $142418328$ & $2428036$ & $5204$ & $0$ & $0$ & $0$\\
$5$ & & $9423021280$ & $9392110424$ & $4793131304$ & $871249112$ & $38702224$ & $304096$ & $264$ & $0$ & $0$ & $0$\\
$6$ & & $8201084228$ & $7129076172$ & $2822253972$ & $334283424$ & $8335124$ & $34644$ & $0$ & $0$ & $0$ & $0$\\
$7$ & & $6588653712$ & $4859106768$ & $1418261432$ & $107410032$ & $1577456$ & $3048$ & $0$ & $0$ & $0$ & $0$\\
$8$ & & $4923371532$ & $2962292832$ & $617683056$ & $28987348$ & $212612$ & $0$ & $0$ & $0$ & $0$ & $0$\\
$9$ & & $3389606120$ & $1599350584$ & $232706440$ & $6459240$ & $22680$ & $0$ & $0$ & $0$ & $0$ & $0$\\
$10$ & & $2159025216$ & $783876948$ & $76440080$ & $1177344$ & $2816$ & $0$ & $0$ & $0$ & $0$ & $0$\\
$11$ & & $1219978864$ & $342570672$ & $21220512$ & $208176$ & $0$ & $0$ & $0$ & $0$ & $0$ & $0$\\
$12$ & & $624337768$ & $133619932$ & $5137464$ & $13904$ & $0$ & $0$ & $0$ & $0$ & $0$ & $0$\\
$13$ & & $272912720$ & $43748496$ & $925824$ & $2304$ & $0$ & $0$ & $0$ & $0$ & $0$ & $0$\\
$14$ & & $108302496$ & $12849536$ & $145120$ & $0$ & $0$ & $0$ & $0$ & $0$ & $0$ & $0$\\
$15$ & & $33279056$ & $2733696$ & $14800$ & $0$ & $0$ & $0$ & $0$ & $0$ & $0$ & $0$\\
$16$ & & $9601392$ & $484528$ & $0$ & $0$ & $0$ & $0$ & $0$ & $0$ & $0$ & $0$\\
$17$ & & $1261536$ & $57984$ & $0$ & $0$ & $0$ & $0$ & $0$ & $0$ & $0$ & $0$\\
$18$ & & $367584$ & $0$ & $0$ & $0$ & $0$ & $0$ & $0$ & $0$ & $0$ & $0$\\
\hline
\hline
\end{tabular}
\end{table}

%% file: list_W_n_r_m_n21to26.tex
\begin{table}[H]
\centering
\caption{\label{tab:W_n_r_m_n21to26}List of $W_{n,m}\big((n-8)a\big)$ for $n=21$--$26$.}
\begin{tabular}{@{}*{10}{r} @{}}
\hline
\hline
\headercell{\\$n$} & \multicolumn{8}{c@{}}{$m$}\\
\cmidrule(l){3-10}
 & & $0$ & $1$ & $2$ & $3$ & $4$ & $5$ & $6$ & $7$\\
\hline
$21$ & & $144322060$ & $147779272$ & $70634012$ & $19763748$ & $3294652$ & $381840$ & $40888$ & $3340$\\
$22$ & & $249379884$ & $242937552$ & $109194420$ & $28422200$ & $4373480$ & $471520$ & $47644$ & $3632$\\
$23$ & & $416838708$ & $387090544$ & $164111960$ & $39902332$ & $5696176$ & $573952$ & $54912$ & $3924$\\
$24$ & & $676458552$ & $599937040$ & $240586012$ & $54847968$ & $7296548$ & $689952$ & $62692$ & $4216$\\
$25$ & & $1069126256$ & $907138184$ & $344982036$ & $73994724$ & $9210556$ & $820336$ & $70984$ & $4508$\\
$26$ & & $1649999040$ & $1341596272$ & $484996292$ & $98176168$ & $11476312$ & $965920$ & $79788$ & $4800$\\
\hline
\hline
\end{tabular}
\end{table}

%% file: list_hat_T_U_inf.tex
\begin{table}[H]
\centering
\caption{\label{tab:hat_T_U_inf_n5to10}List of $\hat{T}_U^\infty(n,r)$ for $n=5$--$10$.}
\begin{tabular}{rrr}
\hline
\hline
$n$ & $r$ & $\hat{T}_U^\infty(n,r)$\\
\hline
$5$ & $3a$ & $-489/2137$\\
\hline
$6$ & $4a$ & $147724/83445$\\
\hline
$7$ & $3a$ & $-797030/993429$\\
$7$ & $5a$ & $3382/2487$\\
\hline
$8$ & $4a$ & $26799212/86839633$\\
$8$ & $6a$ & $7547916/7234927$\\
\hline
$9$ & $3a$ & $-4055477596/5257358209$\\
$9$ & $5a$ & $152081266/83730105$\\
$9$ & $7a$ & $7611583/8876610$\\
\hline
$10$ & $4a$ & $-8701767992/21214395505$\\
$10$ & $6a$ & $40743141476/22363017045$\\
$10$ & $8a$ & $100085804/135876753$\\
\hline
\hline
\end{tabular}
\end{table}

\begin{table}[H]
\centering
\caption{\label{tab:hat_T_U_inf_n11to20}List of $\hat{T}_U^\infty(n,r)$ for $n=11$--$20$.}
\begin{tabular}{rrr}
\hline
\hline
$n$ & $r$ & $\hat{T}_U^\infty(n,r)$\\
\hline
$11$ & $3a$ & $-519108885655/747047679378$\\
$11$ & $5a$ & $150299594390/186104580139$\\
$11$ & $7a$ & $50907107187/33152857465$\\
$11$ & $9a$ & $141364782/217320245$\\
\hline
$12$ & $4a$ & $-107927914267969/214649579344095$\\
$12$ & $6a$ & $105098983928149/57132219401028$\\
$12$ & $8a$ & $1472725414430/1132883874063$\\
$12$ & $10a$ & $142035836/242741103$\\
\hline
$13$ & $3a$ & $-49542736363607596/76900856671970925$\\
$13$ & $5a$ & $2640777994418116/55052813460174941$\\
$13$ & $7a$ & $238175247802385/122518857472329$\\
$13$ & $9a$ & $77953640267/69208879293$\\
$13$ & $11a$ & $406053479/761455803$\\
\hline
$14$ & $4a$ & $-1711574042333777273/4026857160711375875$\\
$14$ & $6a$ & $159995627194416533/141370347617653980$\\
$14$ & $8a$ & $163155707441654/93528121329711$\\
$14$ & $10a$ & $20053990443149/20128254326290$\\
$14$ & $12a$ & $3440746572/7011200365$\\
\hline
$15$ & $3a$ & $-96951623341841992793/150612777410131461334$\\
$15$ & $5a$ & $-691658064075583271/3549165818096600912$\\
$15$ & $7a$ & $70379027228440564/37327326576896655$\\
$15$ & $9a$ & $98381872135352663/64329495834135005$\\
$15$ & $11a$ & $330017198802136/368428513985895$\\
$15$ & $13a$ & $3419428382/7513919679$\\
\hline
$16$ & $4a$ & $-8091159753821109779690/24813727716071921292827$\\
$16$ & $6a$ & $17538926733215035378/36773165966476328825$\\
$16$ & $8a$ & $3298580625991134887/1630566022898346198$\\
$16$ & $10a$ & $46391532219242110/34342463654294913$\\
$16$ & $12a$ & $133661491866764/163885539932895$\\
$16$ & $14a$ & $12880467404/30335604135$\\
\hline
$17$ & $3a$ & $-9215486325414340895326952/13527245403641865728741155$\\
$17$ & $5a$ & $-41809822325295311362381/192527521747950888257130$\\
$17$ & $7a$ & $44666861266339617682797/32415542874022716503150$\\
$17$ & $9a$ & $3527424314659253062003/1862575181539491665133$\\
$17$ & $11a$ & $3426603000355700279/2835203948241401283$\\
$17$ & $13a$ & $38082280733740/50775263002529$\\
$17$ & $15a$ & $1158351747/2909123828$\\
\hline
$18$ & $4a$ & $-76832528089170214024094608/310877703447823092206324125$\\
$18$ & $6a$ & $445196997853554956285398/3167926616427454319937167$\\
$18$ & $8a$ & $22990244319620867204775960/11853824125949561633799469$\\
$18$ & $10a$ & $27085099971224902449460/15824987199150503783871$\\
$18$ & $12a$ & $16484600814607277895/15059491967623573226$\\
$18$ & $14a$ & $11954485092575332/17194567028030343$\\
$18$ & $16a$ & $40040228716/106770751497$\\
\hline
$19$ & $3a$ & $-9988624331188492916121307208/13462710041939624756480714205$\\
$19$ & $5a$ & $-526333026059559368350597412/3183878298414896098809548197$\\
$19$ & $7a$ & $8969653836538056085422828/10733029537802793740799805$\\
$19$ & $9a$ & $10433146285561695510721997/5035720958544970805424924$\\
$19$ & $11a$ & $15253443674321829101249/9896772845670121649875$\\
$19$ & $13a$ & $638884010258507888779/637690480464228768091$\\
$19$ & $15a$ & $16736774506909468/25800961782825345$\\
$19$ & $17a$ & $33417089806/94265509545$\\
\hline
$20$ & $4a$ & $-156424594388085304183446401308/800126304549386100628524904191$\\
$20$ & $6a$ & $5654841561860857134828147888/351604494886093084914378540151$\\
$20$ & $8a$ & $213064242332431581643738876912/136140083511165456790525659591$\\
$20$ & $10a$ & $2060069205296200981486403946/1035875167962083022664474501$\\
$20$ & $12a$ & $23576578880315259956969/16889350416398519119800$\\
$20$ & $14a$ & $17349414652636502449/18755572078860566465$\\
$20$ & $16a$ & $21945750789458799/36063701991448955$\\
$20$ & $18a$ & $108187482924/321802699363$\\
\hline
\hline
\end{tabular}
\end{table}

%% file: 2202_12483.bbl
\begin{thebibliography}{45}%

\bibitem {Staudinger1920}%
    \bibinfo{author}{H.~Staudinger},
    \bibinfo{title}{{\"{U}}ber polymerisation},
    \href{https://doi.org/10.1002/cber.19200530627}{\bibinfo{journal}{Ber. Dtsch. Chem. Ges.}}
    \textbf{\bibinfo{volume}{53}},
    \bibinfo{pages}{1073} 
    (\bibinfo{year}{1920}).

\bibitem {MeyerFerri1935}%
    \bibinfo{author}{K.~H. Meyer}
    and \bibinfo{author}{C.~Ferri},
    \bibinfo{title}{{Sur l'{\'{e}}lasticit{\'{e}} du caoutchouc}},
    \href{https://doi.org/10.1002/hlca.19350180176}{\bibinfo{journal}{Helv. Chim. Acta}}
    \textbf{\bibinfo{volume}{18}},
    \bibinfo{pages}{570}
    (\bibinfo{year}{1935}).

\bibitem{AnthonyGuth1942}%
    \bibinfo{author}{R.~L. Anthony},
    \bibinfo{author}{R.~H. Caston},
    and \bibinfo{author}{E.~Guth},
    \bibinfo{title}{Equations of state for natural and synthetic rubber-like materials. I. Unaccelerated natural soft rubber},
    \href{https://doi.org/10.1021/j150422a005}{\bibinfo{journal}{J. Phys. Chem.}}
    \textbf{\bibinfo{volume}{46}},
    \bibinfo{pages}{826}
    (\bibinfo{year}{1942}).

\bibitem{Mark1976}%
    \bibinfo{author}{J.~E. Mark},
    \bibinfo{title}{Thermoelastic results on rubberlike networks and their bearing on the foundations of elasticity theory},
    \href{https://doi.org/10.1002/POL.1976.230110104}{\bibinfo{journal}{J. Polym. Sci.}}
    \textbf{\bibinfo{volume}{11}},
    \bibinfo{pages}{135}
    (\bibinfo{year}{1976}).

\bibitem{Kittel1969book}%
    \bibinfo{author}{C.~Kittel},
    \textit{\bibinfo{title}{Thermal Physics}}
    (\bibinfo{publisher}{Wiley},
    \bibinfo{address}{New York},
    \bibinfo{year}{1969})
    
\bibitem{Callen1991book}%
    \bibinfo{author}{H.~B. Callen},
    \textit{\bibinfo{title}{{Thermodynamics and an Introduction to Thermostatistics}}}
    (\bibinfo{publisher}{Wiley},
    \bibinfo{address}{New York},
    \bibinfo{year}{1991}).

\bibitem{deGennes1979book}%
    \bibinfo{author}{P.~G. de~Gennes},
    \textit{\bibinfo{title}{Scaling Concepts in Polymer Physics}}
    (\bibinfo{publisher}{Cornell University Press},
    \bibinfo{address}{Ithaca},
    \bibinfo{year}{1979}).

\bibitem{RubinsteinColby2003book}%
    \bibinfo{author}{M.~Rubinstein}
    and \bibinfo{author}{R.~H. Colby},
    \textit{\bibinfo{title}{Polymer Physics}}
    (\bibinfo{publisher}{Oxford University Press},
    \bibinfo{address}{Oxford},
    \bibinfo{year}{2003}).
 
\bibitem{Doi2013book}%
    \bibinfo{author}{M.~Doi},
    \textit{\bibinfo{title}{Soft Matter Physics}}
    (\bibinfo{publisher}{Oxford University Press}, \bibinfo{address}{Oxford}, \bibinfo{year}{2013}).
 
\bibitem{Flory1969book}%
    \bibinfo{author}{P.~J. Flory},
    \textit{\bibinfo{title}{Statistical Mechanics of Chain Molecules}}
    (\bibinfo{publisher}{Wiley Interscience},
    \bibinfo{address}{New York},
    \bibinfo{year}{1969}).
 
\bibitem{YoshikawaSakai2021}%
    \bibinfo{author}{Y.~Yoshikawa},
    \bibinfo{author}{N.~Sakumichi},
    \bibinfo{author}{U.~Chung},
    and \bibinfo{author}{T.~Sakai}, 
    \bibinfo{title}{Negative Energy Elasticity in a Rubberlike Gel},
    \href{https://doi.org/10.1103/PhysRevX.11.011045}{\bibinfo{journal}{Phys. Rev. X}}
    \textbf{\bibinfo{volume}{11}},
    \bibinfo{pages}{011045}
    (\bibinfo{year}{2021}).

\bibitem{SakumichiSakai2021}%
    \bibinfo{author}{N.~Sakumichi},
    \bibinfo{author}{Y.~Yoshikawa},
    and \bibinfo{author}{T.~Sakai},
    \bibinfo{title}{{Linear elasticity of polymer gels in terms of negative energy elasticity}}, 
    \href{https://doi.org/10.1038/s41428-021-00529-4}{\bibinfo{journal}{Polym. J.}}
    \textbf{\bibinfo{volume}{53}},
    \bibinfo{pages}{1293}
    (\bibinfo{year}{2021}).

\bibitem{FujiyabuSakumichi2021}%
    \bibinfo{author}{T.~Fujiyabu},
    \bibinfo{author}{T.~Sakai},
    \bibinfo{author}{R.~Kudo},
    \bibinfo{author}{Y.~Yoshikawa},
    \bibinfo{author}{T.~Katashima},
    \bibinfo{author}{U.~Chung},
    and \bibinfo{author}{N.~Sakumichi},
    \bibinfo{title}{Temperature Dependence of Polymer Network Diffusion},
    \href{https://doi.org/10.1103/PhysRevLett.127.237801}{\bibinfo{journal}{Phys. Rev. Lett.}}
    \textbf{\bibinfo{volume}{127}},
    \bibinfo{pages}{237801}
    (\bibinfo{year}{2021}).
 
\bibitem{Orr1947}%
    \bibinfo{author}{W.~J. Orr},
    \bibinfo{title}{Statistical treatment of polymer solutions at infinite dilution},
    \href{https://doi.org/10.1039/tf9474300012}{\bibinfo{journal}{Trans. Faraday Soc.}}
    \textbf{\bibinfo{volume}{43}},
    \bibinfo{pages}{12}
    (\bibinfo{year}{1947}).
 
\bibitem{MadrasSlade1996book}%
    \bibinfo{author}{N.~Madras}
    and \bibinfo{author}{G.~Slade},
    \href{https://doi.org/10.1007/978-1-4612-4132-4}{\textit{\bibinfo{title}{The Self-Avoiding Walk}}},
    Probability and Its Applications
    (\bibinfo{publisher}{Birkh{\"{a}}user},
    \bibinfo{address}{Boston},
    \bibinfo{year}{1996}).
 
\bibitem{BroederszMacKintosh2014}%
    \bibinfo{author}{C.~P. Broedersz}
    and \bibinfo{author}{F.~C. MacKintosh},
    \bibinfo{title}{Modeling semiflexible polymer networks},
    \href{https://doi.org/10.1103/RevModPhys.86.995}{\bibinfo{journal}{Rev. Mod. Phys.}}\textbf{\bibinfo{volume}{86}},
    \bibinfo{pages}{995}
    (\bibinfo{year}{2014}).
 
\bibitem{Vanderzande1998book}%
    \bibinfo{author}{C.~Vanderzande},
    \href{https://doi.org/10.1017/CBO9780511563935}{\textit{\bibinfo{title}{Lattice Models of Polymers}}},
    Cambridge Lecture Notes in Physics
    (\bibinfo{publisher}{Cambridge University Press},
    \bibinfo{address}{Cambridge},
    \bibinfo{address}{England},
    \bibinfo{year}{1998}).
 
\bibitem{Rapaport1974}%
    \bibinfo{author}{D.~C. Rapaport},
    \bibinfo{title}{{On the polymer phase transition}},
    \href{https://doi.org/10.1016/0375-9601(74)90456-3}{\bibinfo{journal}{Phys. Lett.}}
    \textbf{\bibinfo{volume}{48A}},
    \bibinfo{pages}{339}
    (\bibinfo{year}{1974}).
 
\bibitem{MarenduzzoSeno2003}%
    \bibinfo{author}{D.~Marenduzzo},
    \bibinfo{author}{A.~Maritan},
    \bibinfo{author}{A.~Rosa},
    and
    \bibinfo{author}{F.~Seno},
    \bibinfo{title}{Stretching of a Polymer Below the $\theta$ Point},
    \href{https://doi.org/10.1103/PhysRevLett.90.088301}{\bibinfo{journal}{Phys. Rev. Lett.}}
    \textbf{\bibinfo{volume}{90}},
    \bibinfo{pages}{088301}
    (\bibinfo{year}{2003}).
 
\bibitem{KumarGuttmann2007}%
    \bibinfo{author}{S.~Kumar},
    \bibinfo{author}{I.~Jensen},
    \bibinfo{author}{J.~L. Jacobsen},
    and \bibinfo{author}{A.~J. Guttmann},\
    \bibinfo{title}{Role of Conformational Entropy in Force-Induced Biopolymer Unfolding},
    \href{https://doi.org/10.1103/PhysRevLett.98.128101}{\bibinfo{journal}{Phys. Rev. Lett.}}
    \textbf{\bibinfo{volume}{98}},
    \bibinfo{pages}{128101}
    (\bibinfo{year}{2007}).

\bibitem{LeeLee2012}%
    \bibinfo{author}{J.~H. Lee},
    \bibinfo{author}{S.~Y. Kim},
    and \bibinfo{author}{J.~Lee},
    \bibinfo{title}{Exact partition function zeros of a polymer on a simple cubic lattice},
    \href{https://doi.org/10.1103/PhysRevE.86.011802}
    {\bibinfo{journal}{Phys. Rev. E}}
    \textbf{\bibinfo{volume} {86}},
    \bibinfo{pages}{011802} (\bibinfo{year}{2012}).
 
\bibitem{HsiehHu2016aug}%
    \bibinfo{author}{Y.~H. Hsieh},
    \bibinfo{author}{C.~N. Chen},
    and
    \bibinfo{author}{C.~K. Hu},
    \bibinfo{title}{Efficient algorithm for computing exact partition functions of lattice polymer models},
    \href{https://doi.org/10.1016/J.CPC.2016.08.006}{\bibinfo{journal}{Comput. Phys. Commun.}}
    \textbf{\bibinfo{volume}{209}},
    \bibinfo{pages}{27}
    (\bibinfo{year}{2016}).
 
\bibitem{SMcommentGelSAW2021} See Supplemental Material for (i) details of the enumeration method, (ii) derivations of analytic expressions, (iii) discussion on the data collapse shown in Fig.~\ref{fig:T_U_inf_n20}(b), and (iv) complete lists of enumeration data, which includes Refs.~\cite{SchramBisseling2011,SchramBisseling2013,ClisbySlade2007}. 

\bibitem{SchramBisseling2011}%
    \bibinfo{author}{R.~D. Schram},
    \bibinfo{author}{G.~T. Barkema},
    and
    \bibinfo{author}{R.~H. Bisseling},
    \bibinfo{title}{Exact enumeration of self-avoiding walks},
    \href{https://doi.org/10.1088/1742-5468/2011/06/p06019}{\bibinfo{journal}{J. Stat. Mech.}}
    (\bibinfo{year}{2011})
    \bibinfo{pages}{P06019}.
 
\bibitem{SchramBisseling2013}%
    \bibinfo{author}{R.~D. Schram},
    \bibinfo{author}{G.~T. Barkema},
    and
    \bibinfo{author}{R.~H. Bisseling},
    \bibinfo{title}{SAWdoubler: A program for counting self-avoiding walks},
    \href{https://doi.org/10.1016/j.cpc.2012.10.026}{\bibinfo{journal}{Comput. Phys. Commun.}}
    \textbf{\bibinfo{volume}{184}},
    \bibinfo{pages}{891}
    (\bibinfo{year}{2013}).

\bibitem{ClisbySlade2007}%
    \bibinfo{author}{N.~Clisby},
    \bibinfo{author}{R.~Liang},
    and
    \bibinfo{author}{G.~Slade},
    \bibinfo{title}{Self-avoiding walk enumeration via the lace
    expansion}, 
    \href{https://doi.org/10.1088/1751-8113/40/36/003}{\bibinfo{journal}{J. Phys. A}}
    \textbf{\bibinfo{volume} {40}},
    \bibinfo{pages}{10973}
    (\bibinfo{year} {2007}).

\bibitem{Martin1962}%
    \bibinfo{author}{J.~L. Martin}, 
    \bibinfo{title}{{The exact enumeration of self-avoiding walks on a lattice}},
    \href{https://doi.org/10.1017/S0305004100036240}{\bibinfo{journal}{Math. Proc. Cambridge Philos. Soc.}}
    \textbf{\bibinfo{volume} {58}},
    \bibinfo{pages}{92}
    (\bibinfo{year}{1962}).

\bibitem{FisherSykes1959}%
    \bibinfo{author}{M.~E. Fisher}
    and \bibinfo{author}{M.~F. Sykes},
    \bibinfo{title}{Excluded-volume problem and the Ising model of ferromagnetism},
    \href{https://doi.org/10.1103/PhysRev.114.45}{\bibinfo{journal}{Phys. Rev.}}
    \textbf{\bibinfo{volume}{114}},
    \bibinfo{pages}{45}
    (\bibinfo{year}{1959}).
 
\bibitem{Sykes1961}%
      \bibinfo{author}{M.~F. Sykes},
      \bibinfo{title}{Some counting theorems in the theory of the Ising model and the excluded volume problem},
      \href{https://doi.org/10.1063/1.1724212}{\bibinfo{journal}{J. Math. Phys.}}
      \textbf{\bibinfo{volume}{2}},
       \bibinfo{pages}{52}
       (\bibinfo{year}{1961}).
 
\bibitem{Sykes1963}%
    \bibinfo{author}{M.~F. Sykes},
    \bibinfo{title}{{Self‐avoiding walks on the simple cubic lattice}},
    \href{https://doi.org/10.1063/1.1734262}{\bibinfo{journal}{J. Chem. Phys.}}
    \textbf{\bibinfo{volume}{39}},
    \bibinfo{pages}{410}
    (\bibinfo{year}{1963}).
 
\bibitem{FisherHiley1961}%
    \bibinfo{author}{M.~E. Fisher}
    and \bibinfo{author}{B.~J. Hiley},
    \bibinfo{title}{Configuration and free energy of a polymer molecule with solvent interaction},
    \href{https://doi.org/10.1063/1.1731729}{\bibinfo{journal} {J. Chem. Phys.}}
    \textbf{\bibinfo{volume}{34}},
    \bibinfo{pages}{1253}
    (\bibinfo{year}{1961}).
 
\bibitem{NemirovskyDouglas1992}%
    \bibinfo{author}{A.~M. Nemirovsky},
    \bibinfo{author}{K.~F. Freed},
    \bibinfo{author}{T.~Ishinabe},
    and \bibinfo{author}{J.~F. Douglas},
    \bibinfo{title}{Marriage of exact enumeration and 1/d expansion methods: Lattice model of dilute polymers},
    \href{https://doi.org/10.1007/BF01049010}{\bibinfo{journal}{J. Stat. Phys.}}
    \textbf{\bibinfo{volume}{67}},
    \bibinfo{pages}{1083}
    (\bibinfo{year}{1992}).
 
\bibitem{ButeraComi1999}%
    \bibinfo{author}{P.~Butera}
    and \bibinfo{author}{M.~Comi},
    \bibinfo{title}{Critical specific heats of the N-vector spin models on the simple cubic and bcc lattices},
    \href{https://doi.org/10.1103/PhysRevB.60.6749}{\bibinfo{journal}{Phys. Rev. B}}
    \textbf{\bibinfo{volume}{60}},
    \bibinfo{pages} {6749}
    (\bibinfo{year}{1999}).
 
\bibitem{Guttmann1987}%
    \bibinfo{author}{A.~J. Guttmann},
    \bibinfo{title}{{On the critical behaviour of self-avoiding walks}},
    \href{https://doi.org/10.1088/0305-4470/20/7/029}{\bibinfo{journal}{J. Phys. A}}
    \textbf{\bibinfo{volume}{20}},
    \bibinfo{pages}{1839}
    (\bibinfo{year}{1987}).
 
\bibitem{DombWilmers1965}%
    \bibinfo{author}{C.~Domb},
    \bibinfo{author}{J.~Gillis},
    and \bibinfo{author}{G.~Wilmers},
    \bibinfo{title}{{On the shape and configuration of polymer molecules}},
    \href{https://doi.org/10.1088/0370-1328/85/4/302}{\bibinfo{journal}{Proc. Phys. Soc. London}}
    \textbf{\bibinfo {volume}{85}},
    \bibinfo{pages}{625}
    (\bibinfo{year}{1965}).
 
\bibitem{A001412}%
    \bibinfo{author}{N.~J.~A. Sloane},
    \bibinfo{title}{Entry A001412 in The On-Line Encyclopedia of Integer Sequences, \url{https://oeis.org/A001412}}
    (\bibinfo{year}{2022}).
 
\bibitem{A174319}%
    \bibinfo{author}{J.~Myers},
    \bibinfo{title}{Entry A174319 in The On-Line Encyclopedia of Integer Sequences, \url{https://oeis.org/A174319}}
    (\bibinfo{year}{2022}).
 
\bibitem{Nienhuis1982}%
    \bibinfo{author}{B.~Nienhuis},
    \bibinfo{title}{Exact Critical Point and Critical Exponents of $\mathrm{O}(n)$ Models in Two Dimensions},
    \href{https://doi.org/10.1103/PhysRevLett.49.1062}{\bibinfo{journal}{Phys. Rev. Lett.}}
    \textbf{\bibinfo{volume}{49}},
    \bibinfo{pages}{1062}
    (\bibinfo{year}{1982}).
 
\bibitem{Nienhuis1984}%
    \bibinfo{author}{B.~Nienhuis},
    \bibinfo{title}{Critical behavior of two-dimensional spin models and charge asymmetry in the Coulomb gas},
    \href{https://doi.org/10.1007/BF01009437}{\bibinfo{journal}{J. Stat. Phys.}}
    \textbf{\bibinfo{volume}{34}},
    \bibinfo{pages}{731}
    (\bibinfo{year}{1984}).
 
\bibitem{TorrieWhittington1975}%
    \bibinfo{author}{G.~Torrie}
    and \bibinfo{author}{S.~G. Whittington},
    \bibinfo{title}{Exact enumeration of neighbour-avoiding walks on the tetrahedral and body-centred cubic lattices},
    \href{https://doi.org/10.1088/0305-4470/8/7/022}{\bibinfo{journal}{J. Phys. A}}
    \textbf{\bibinfo{volume}{8}},
    \bibinfo{pages}{1178}
    (\bibinfo{year}{1975}).

\bibitem{IshinabeChikahisa1986}%
    \bibinfo{author}{T.~Ishinabe}
    and \bibinfo{author}{Y.~Chikahisa},
    \bibinfo{title}{Exact enumerations of self‐avoiding lattice walks with different nearest‐neighbor contacts},
    \href{https://doi.org/10.1063/1.451345}{\bibinfo{journal}{J. Chem. Phys.}}
    \textbf{\bibinfo{volume}{85}},
    \bibinfo{pages}{1009}
    (\bibinfo{year}{1986}).

\bibitem{NakajimaNishi2006}%
    \bibinfo{author}{K.~Nakajima},
    \bibinfo{author}{H.~Watabe},
    and \bibinfo{author}{T.~Nishi},
    \bibinfo{title}{Single polymer chain rubber elasticity investigated by atomic force microscopy},
    \href{https://doi.org/10.1016/j.polymer.2005.12.092}{\bibinfo{journal}{Polymer}}
    \textbf{\bibinfo{volume}{47}},
    \bibinfo{pages}{2505}
    (\bibinfo{year}{2006}).
 
\bibitem{KolbergBalzer2019}%
    \bibinfo{author}{A.~Kolberg},
    \bibinfo{author}{C.~Wenzel},
    \bibinfo{author}{K.~Hackenstrass},
    \bibinfo{author}{R.~Schwarzl},
    \bibinfo{author}{C.~R{\"{u}}ttiger},
    \bibinfo{author}{T.~Hugel},
    \bibinfo{author}{M.~Gallei},
    \bibinfo{author}{R.~R. Netz},
    and \bibinfo{author}{B.~N. Balzer},
    \bibinfo{title}{Opposing temperature dependence of the stretching response of single PEG and PNiPAM polymers},
    \href{https://doi.org/10.1021/jacs.9b04383}{\bibinfo{journal}{J. Am. Chem. Soc.}}
    \textbf{\bibinfo{volume}{141}},
    \bibinfo{pages}{11603}
    (\bibinfo{year}{2019}).
 
\bibitem{BaoCui2020}%
  \bibinfo{author}{Y.~Bao},
  \bibinfo{author}{Z.~Luo},
  and  \bibinfo{author}{S.~Cui},
  \bibinfo{title}{Environment-dependent single-chain mechanics of synthetic polymers and biomacromolecules by atomic force microscopy-based single-molecule force spectroscopy and the implications for advanced polymer materials},
  \href{https://doi.org/10.1039/C9CS00855A}{\bibinfo{journal}{Chem. Soc. Rev.}}
  \textbf{\bibinfo{volume}{49}},
  \bibinfo{pages}{2799}
  (\bibinfo{year}{2020}).

\end{thebibliography}
